\documentclass[a4paper,11pt]{article}
\pdfoutput=1 

\usepackage{jheppub} 

\usepackage[T1]{fontenc} 
\usepackage{subfig}
\usepackage{enumerate}

\makeatletter
\DeclareRobustCommand*{\bfseries}{%
   \not@math@alphabet\bfseries\mathbf
   \fontseries\bfdefault\selectfont
   \boldmath
}
\makeatother

\usepackage{wasysym}

\usepackage{bm}
\usepackage{tikz}
\usepackage{tikz-3dplot}
\usepackage{verbatim}
\usetikzlibrary{decorations.pathreplacing}
\usepackage{mathrsfs}

\usetikzlibrary{decorations.pathreplacing,decorations.markings}

\tikzset{
    >=stealth',
    punkt/.style={
           rectangle,
           rounded corners,
           draw=black, very thick,
           text width=6.5em,
           minimum height=2em,
           text centered},
    pil/.style={
           ->,
           thick,
           shorten <=2pt,
           shorten >=2pt,},
  on each segment/.style={
    decorate,
    decoration={
      show path construction,
      moveto code={},
      lineto code={
        \path [#1]
        (\tikzinputsegmentfirst) -- (\tikzinputsegmentlast);
      },
      curveto code={
        \path [#1] (\tikzinputsegmentfirst)
        .. controls
        (\tikzinputsegmentsupporta) and (\tikzinputsegmentsupportb)
        ..
        (\tikzinputsegmentlast);
      },
      closepath code={
        \path [#1]
        (\tikzinputsegmentfirst) -- (\tikzinputsegmentlast);
      },
    },
  },
  mid arrow/.style={postaction={decorate,decoration={
        markings,
        mark=at position .5 with {\arrow[#1]{stealth'}}
      }}}
}

 \newcommand{\ket}[1]{|#1\rangle}

\newtheorem{theorem}{Theorem}

\newtheorem{definition}[theorem]{Definition}

\newtheorem{lemma}[theorem]{Lemma}

\newenvironment{argument}[1][Argument]{\noindent\textbf{#1.} }{\ \rule{0.5em}{0.5em}}

\usetikzlibrary{decorations.pathmorphing}
\tikzset{snake it/.style={decorate, decoration=snake}}

\title{Holographic quantum tasks with input and output regions}

\author[a]{Alex May}

\affiliation[a]{Department of Physics and Astronomy, University of British Columbia
6224 Agricultural Road, Vancouver, B.C., V6T 1W9, Canada}

\emailAdd{may@phas.ubc.ca}

\abstract{Quantum tasks are quantum computations with inputs and outputs occurring at specified spacetime locations. Considering such tasks in the context of AdS/CFT has led to novel constraints relating bulk geometry and boundary entanglement. In this article we consider tasks where inputs and outputs are encoded into extended spacetime regions, rather than the points previously considered. We show that this leads to stronger constraints than have been derived in the point based setting. In particular we improve the connected wedge theorem, appearing earlier in 1912.05649, by finding a larger bulk region whose existence implies large boundary correlation. As well, we show how considering extended input and output regions leads to non-trivial statements in Poincar\'e-AdS$_{2+1}$, a setting where the point-based connected wedge theorem is always trivial.}

\begin{document} 

\maketitle


\section{Introduction}

\begin{figure}
    \centering
    \subfloat[\label{fig:introconnected-scattering}]{
    \tdplotsetmaincoords{15}{0}
    \begin{tikzpicture}[scale=1.4,tdplot_main_coords]
    
    \draw[gray] (-2,1,0) -- (-2,5,0);
    \draw[gray] (2,1,0) -- (2,5,0);
    \tdplotsetrotatedcoords{0}{45}{0}
    \begin{scope}[tdplot_rotated_coords]
    
    
    \draw[thick,red,-triangle 45] (0,1.33,-1.55) -- (0,3,0);
    \draw[red] plot [mark=*, mark size=1.5] coordinates{(0,1.33,-1.55)};
    \node[above right] at (0.1,1.33,-1.55) {$c_1$};
    
    \draw[domain=0:45,variable=\x,smooth, fill=purple,opacity=0.6] plot ({-2*sin(\x)}, {1+\x/45}, {2*cos(\x)}) -- plot ({-2*sin((45-\x))}, {3-(45-\x)/45}, {2*cos(45-\x)}) --  plot ({2*sin(\x)}, {3-\x/45}, {2*cos(\x)}) -- plot ({2*sin(45-\x)}, {1+(45-\x)/45}, {2*cos(45-\x)});
    
    \draw[domain=0:45,variable=\x,smooth,thick] plot ({-2*sin(\x)}, {1+\x/45}, {2*cos(\x)});
    \draw[domain=0:45,variable=\x,smooth,thick] plot ({2*sin(\x)}, {1+\x/45}, {2*cos(\x)});
    \draw[domain=0:45,variable=\x,smooth,thick] plot ({-2*sin(\x)}, {3-\x/45}, {2*cos(\x)});
    \draw[domain=0:45,variable=\x,smooth,thick] plot ({2*sin(\x)}, {3-\x/45}, {2*cos(\x)});
    
    \begin{scope}[canvas is xz plane at y=1]
    \draw[gray] (0,0) circle[radius=2] ;
    \end{scope}
    
    \begin{scope}[canvas is xz plane at y=5]
    \draw[gray] (0,0) circle[radius=2] ;
    \end{scope}
    
    \draw [domain=-45:45] plot ({2*cos(\x+90)},2, {2*sin(\x+90)});
    
    
    
    \draw[thick,red,-triangle 45] (0,2,0.9) -- (0,3,0);
    
    \foreach \x in {-16,...,-164}
    {
    \draw[black,opacity=0.3] (0,1,-2) -- ({0.55*cos(\x)},{1.33},{-0.55*sin(\x)-2.1});
    \draw[black,opacity=0.3] (0,1.66,-2) -- ({0.55*cos(\x)},{1.33},{-0.55*sin(\x)-2.1});
    }
    
    \draw[domain=-15:15,variable=\x,smooth,thick] plot ({2*sin(\x+180)}, {1.333}, {2*cos(\x+180)});
    \draw[domain=0:15,variable=\x,smooth, thick] plot ({2*sin(\x+180)}, {1+\x/45}, {2*cos(\x)});
    \draw[domain=0:-15,variable=\x,smooth, thick] plot ({2*sin(\x+180)}, {1-\x/45}, {2*cos(\x+180)});
    \draw[domain=0:-15,variable=\x,smooth, thick] plot ({2*sin(\x+180)}, {1.666+\x/45}, {2*cos(\x+180)});
    \draw[domain=0:15,variable=\x,smooth, thick] plot ({2*sin(\x+180)}, {1.666-\x/45}, {2*cos(\x+180)});
    
    \draw[domain=-16:-164, thick] plot ({0.55*cos(\x)},{1.33},{-0.55*sin(\x)-2.1});
    
    \draw[domain=0:45,variable=\x,smooth, fill=purple,opacity=0.6] plot ({-2*sin(\x+180)}, {1+\x/45}, {2*cos(\x+180)}) -- plot ({-2*sin((45-\x)+180)}, {3-(45-\x)/45}, {2*cos(45-\x+180)}) --  plot ({2*sin(\x+180)}, {3-\x/45}, {2*cos(\x+180)}) -- plot ({2*sin(45-\x+180)}, {1+(45-\x)/45}, {2*cos(45-\x+180)});
    
    \begin{scope}[canvas is xz plane at y=2]
    \draw [domain=-45:45] plot ({2*cos(\x-90)}, {2*sin(\x-90)});
    \end{scope}
    
    \draw[domain=0:45,variable=\x,smooth,thick] plot ({-2*sin(\x+180)}, {1+\x/45}, {2*cos(\x+180)});
    
    \draw[domain=0:45,variable=\x,smooth,thick] plot ({2*sin(\x+180)}, {1+\x/45}, {2*cos(\x+180)});
    
    \draw[domain=0:45,variable=\x,smooth,thick] plot ({-2*sin(\x+180)}, {3-\x/45}, {2*cos(\x+180)});
    
    \draw[domain=0:45,variable=\x,smooth,thick] plot ({2*sin(\x+180)}, {3-\x/45}, {2*cos(\x+180)});

    \draw[red] plot [mark=*, mark size=1.5] coordinates{(0,3,0)};
    
    \node at (-2,1.4,0) {$\hat{\mathcal{V}}_1$};
    \node at (-2.6,1.4,4) {$\hat{\mathcal{V}}_2$};
    
    \draw[red] plot [mark=*, mark size=1.5] coordinates{(0.9,4,0)};
    \node[below right] at (0.9,4.1,0) {$r_2$};
    
    \draw[red,-triangle 45, thick] (0,3,0)--(0.9,4,0);
    
    \draw[domain=0:90,variable=\x,smooth,dashed] plot ({-2*sin(\x+180)}, {3+\x/45}, {2*cos(\x+180)});
    \draw[domain=0:90,variable=\x,smooth,dashed] plot ({-2*sin(\x+180)}, {3+\x/45}, {-2*cos(\x+180)});
    
    \draw[domain=0:90,variable=\x,smooth,dashed] plot ({2*sin(\x+180)}, {3+\x/45}, {2*cos(\x+180)});
    \draw[domain=0:90,variable=\x,smooth,dashed] plot ({2*sin(\x+180)}, {3+\x/45}, {-2*cos(\x+180)});
    
    \draw[domain=0:45,variable=\x,smooth, fill=blue,opacity=0.3] plot ({-2*sin(\x+90)}, {3+\x/45}, {2*cos(\x+90)}) -- plot ({-2*sin((45-\x+90))}, {5-(45-\x)/45}, {2*cos(45-\x+90)}) --  plot ({2*sin(\x-90)}, {5-\x/45}, {2*cos(\x-90)}) -- plot ({2*sin(45-\x-90)}, {3+(45-\x)/45}, {2*cos(45-\x-90)});
    
    \draw[domain=0:45,variable=\x,smooth,thick,blue] plot ({-2*sin(\x+90)}, {3+\x/45}, {2*cos(\x+90)});
    \draw[domain=0:45,variable=\x,smooth,thick,blue] plot ({2*sin(\x-90)}, {3+\x/45}, {2*cos(\x-90)});
    \draw[domain=0:45,variable=\x,smooth,thick,blue] plot ({-2*sin(\x+90)}, {5-\x/45}, {2*cos(\x+90)});
    \draw[domain=0:45,variable=\x,smooth,thick,blue] plot ({2*sin(\x-90)}, {5-\x/45}, {2*cos(\x-90)});

    \draw [domain=-135:-45,blue] plot ({2*cos(\x-90)},4,{2*sin(\x-90)});
    \draw [domain=135:45,blue] plot ({2*cos(\x-90)},4,{2*sin(\x-90)});
    
    \draw[domain=0:45,variable=\x,smooth, fill=blue,opacity=0.3] plot ({-2*sin(\x-90)}, {3+\x/45}, {2*cos(\x-90)}) -- plot ({-2*sin((45-\x-90))}, {5-(45-\x)/45}, {2*cos(45-\x-90)}) --  plot ({2*sin(\x+90)}, {5-\x/45}, {2*cos(\x+90)}) -- plot ({2*sin(45-\x+90)}, {3+(45-\x)/45}, {2*cos(45-\x+90)});
    
    \draw[domain=0:45,variable=\x,smooth,thick,blue] plot ({-2*sin(\x-90)}, {3+\x/45}, {2*cos(\x-90)});
    \draw[domain=0:45,variable=\x,smooth,thick,blue] plot ({2*sin(\x+90)}, {3+\x/45}, {2*cos(\x+90)});
    \draw[domain=0:45,variable=\x,smooth,thick,blue] plot ({-2*sin(\x-90)}, {5-\x/45}, {2*cos(\x-90)});
    \draw[domain=0:45,variable=\x,smooth,thick,blue] plot ({2*sin(\x+90)}, {5-\x/45}, {2*cos(\x+90)});
    
    \draw[domain=45:135, thick, blue] plot
    ({2*cos(\x+90)+2.85},{4},{2*sin(\x+90)});
    \draw[domain=45:135, thick, blue] plot
    ({2*cos(\x-90)-2.85},{4},{-2*sin(\x-90)});
    
    \foreach \x in {45,...,135}
    {
    \draw[opacity=0.3,blue] (2,5,0) -- ({2*cos(\x+90)+2.85},{4},{2*sin(\x+90)});
    \draw[opacity=0.3,blue] (-2,5,0) -- ({2*cos(\x-90)-2.85},{4},{-2*sin(\x-90)});
    }
    
    \foreach \x in {45,...,135}
    {
    \draw[opacity=0.3,blue] (2,3,0) -- ({2*cos(\x+90)+2.85},{4},{2*sin(\x+90)});
    \draw[opacity=0.3,blue] (-2,3,0) -- ({2*cos(\x-90)-2.85},{4},{-2*sin(\x-90)});
    }
    
    \foreach \x in {11,14,...,41}
    {
    \draw[opacity=0.3,blue] plot (2,3,0) -- ({2*sin(\x+90)}, {3+(\x)/45}, {2*cos(\x+90)});
    }

    \foreach \x in {11,14,...,41}
    {
    \draw[opacity=0.3,blue] plot (-2,3,0) -- ({2*sin(\x-90)}, {3+(\x)/45}, {2*cos(\x-90)});
    }
    
    \foreach \x in {11,14,...,41}
    {
    \draw[opacity=0.3,blue] plot (-2,3,0) -- ({2*sin(\x-90)}, {3+(\x)/45}, {-2*cos(\x-90)});
    }

    \foreach \x in {11,14,...,41}
    {
    \draw[opacity=0.3,blue] plot (2,3,0) -- ({2*sin(\x+90)}, {3+(\x)/45}, {-2*cos(\x+90)});
    }
    
    \foreach \x in {11,14,...,41}
    {
    \draw[opacity=0.3,blue] plot (2,5,0) -- ({2*sin(\x+90)}, {5-(\x)/45}, {2*cos(\x+90)});
    }

    \foreach \x in {11,14,...,41}
    {
    \draw[opacity=0.3,blue] plot (-2,5,0) -- ({2*sin(\x-90)}, {5-(\x)/45}, {2*cos(\x-90)});
    }
    
    \foreach \x in {11,14,...,41}
    {
    \draw[opacity=0.3,blue] plot (-2,5,0) -- ({2*sin(\x-90)}, {5-(\x)/45}, {-2*cos(\x-90)});
    }

    \foreach \x in {11,14,...,41}
    {
    \draw[opacity=0.3,blue] plot (2,5,0) -- ({2*sin(\x+90)}, {5-(\x)/45}, {-2*cos(\x+90)});
    }

    \draw[red] plot [mark=*, mark size=1.5] coordinates{(-0.9,4,0)};
    \node[above] at (-0.9,4,0) {$r_1$};
    \draw[red,-triangle 45,thick] (0,3,0)--(-0.9,4,0);
    
    \node at (-2,4.2,0) {${\mathcal{R}_1}$};
    \node at (-2,2.7,4) {${\mathcal{R}_2}$};
    
    \foreach \x in {16,...,164}
    {
    \draw[black,opacity=0.3] (0,1,2) -- ({0.55*cos(\x)},{1.33},{-0.55*sin(\x)+2.1});
    \draw[black,opacity=0.3] (0,1.66,2) -- ({0.55*cos(\x)},{1.33},{-0.55*sin(\x)+2.1});
    }
    
    \draw[domain=-15:15,variable=\x,smooth,thick] plot ({2*sin(\x)}, {1.333}, {2*cos(\x)});
    \draw[domain=0:15,variable=\x,smooth, thick] plot ({2*sin(\x)}, {1+\x/45}, {2*cos(\x)});
    \draw[domain=0:-15,variable=\x,smooth, thick] plot ({2*sin(\x)}, {1-\x/45}, {2*cos(\x)});
    \draw[domain=0:-15,variable=\x,smooth, thick] plot ({2*sin(\x)}, {1.666+\x/45}, {2*cos(\x)});
    \draw[domain=0:15,variable=\x,smooth, thick] plot ({2*sin(\x)}, {1.666-\x/45}, {2*cos(\x)});
    
    \draw[domain=16:164, thick] plot ({0.55*cos(\x)},{1.33},{-0.55*sin(\x)+2.1});
    
    \draw[thick,red,-triangle 45] (0,1.33,1.5) -- (0,3,0);
    \draw[red] plot [mark=*, mark size=1.5] coordinates{(0,1.33,1.5)};
    \node[below left] at (0,1.33,1.5) {$c_2$};
    
    \end{scope}
    \end{tikzpicture}
    }
    \hfill
    \centering
    \subfloat[\label{fig:boundaryregionsintro}]{
    \begin{tikzpicture}[scale=1.6]

    \draw (-2,0) -- (2,0) -- (2,2) -- (-2,2) -- (-2,0);
    
    \draw[fill=purple,opacity=0.6] (0,0) -- (0.5,0.5) -- (0,1) -- (-0.5,0.5) -- (0,0);
    \draw[thick] (0,0) -- (0.5,0.5) -- (0,1) -- (-0.5,0.5) -- (0,0);
    \node at (0,0.6) {$\hat{\mathcal{V}}_1$};
    
    \draw[fill=purple,opacity=0.6] (-2,0) -- (-1.5,0.5) -- (-2,1) -- (-2,0);
    \draw[thick] (-2,0) -- (-1.5,0.5) -- (-2,1) -- (-2,0);
    \draw[fill=purple,opacity=0.6] (2,0) -- (1.5,0.5) -- (2,1) -- (2,0);
    \draw[thick] (2,0) -- (1.5,0.5) -- (2,1) -- (2,0);
    
    \node at (1.82,0.6) {$\hat{\mathcal{V}}_2$};
    \node at (-1.82,0.6) {$\hat{\mathcal{V}}_2$};
    
    \draw[fill=black,opacity=0.6] (0,0) -- (0.2,0.2)--(0,0.4) -- (-0.2,0.2)--(0,0);
    
    \draw[fill=black,opacity=0.6] (-2,0) -- (-1.8,0.2) -- (-2,0.4);
    \draw[fill=black,opacity=0.6] (2,0) -- (1.8,0.2) -- (2,0.4);
    
    \node at (-0.2,-0.2) {$\hat{\mathcal{C}}_1$};
    \draw[->] (-0.2,-0.07) to [out=90,in=-160] (0,0.2);
    
    \node at (-2.2,-0.2) {$\hat{\mathcal{C}}_2$};
    \draw[->] (-2.2,-0.07) to [out=90,in=-160] (-2,0.2);
    
    \draw[fill=blue,opacity=0.3] (-1,2) -- (-1.3,1.7) -- (-1,1.4) -- (-0.7,1.7) -- (-1,2);
    \draw[thick,blue] (-1,2) -- (-1.3,1.7) -- (-1,1.4) -- (-0.7,1.7) -- (-1,2);
    
    \node at (-1,1.7) {$\hat{\mathcal{R}}_1$};
    
    \draw[fill=blue,opacity=0.3] (1,2) -- (1.3,1.7) -- (1,1.4) -- (0.7,1.7) -- (1,2);
    \draw[thick,blue] (1,2) -- (1.3,1.7) -- (1,1.4) -- (0.7,1.7) -- (1,2);
    
    \node at (1,1.7) {$\hat{\mathcal{R}}_2$};
    
    \node at (0,-1) {$ $};
    
    \draw[dashed] (0.7,1.7) -- (0,1);
    \draw[dashed] (-0.7,1.7) -- (0,1);
    
    \draw[dashed] (1.3,1.7) -- (2,1);
    \draw[dashed] (-1.3,1.7) -- (-2,1);
    
    \end{tikzpicture}
    }
    \caption{(a) Bulk perspective on a particular quantum task. The task has inputs given at bulk points $c_1$ and $c_2$, and outputs are required at $r_1$ and $r_2$. (b) Boundary view of the same task. Now regions $\hat{\mathcal{C}}_i$ whose entanglement wedge $\mathcal{C}_i$ contains point $c_i$ become the input regions, while the $\hat{\mathcal{R}}_i$ whose entanglement wedge $\hat{\mathcal{R}}_i$ contains the $r_i$ become the output regions. Define the scattering region $J^E_{12\rightarrow 12}\equiv J^+(\mathcal{C}_1)\cap J^+(\mathcal{C}_2) \cap J^-(\mathcal{R}_1) \cap J^-(\mathcal{R}_2)$ in the bulk geometry, and the decision regions $\hat{\mathcal{V}}_i\equiv \hat{J}^+(\hat{\mathcal{C}}_i)\cap \hat{J}^-(\hat{\mathcal{R}}_1)\cap \hat{J}^-(\hat{\mathcal{R}}_2)$. The connected wedge theorem states that when $J^E_{12\rightarrow 12}$ is non-empty, the entanglement wedge of $\hat{\mathcal{V}}_1\cup \hat{\mathcal{V}}_2$ is connected. }
    \label{fig:introfig}
\end{figure}
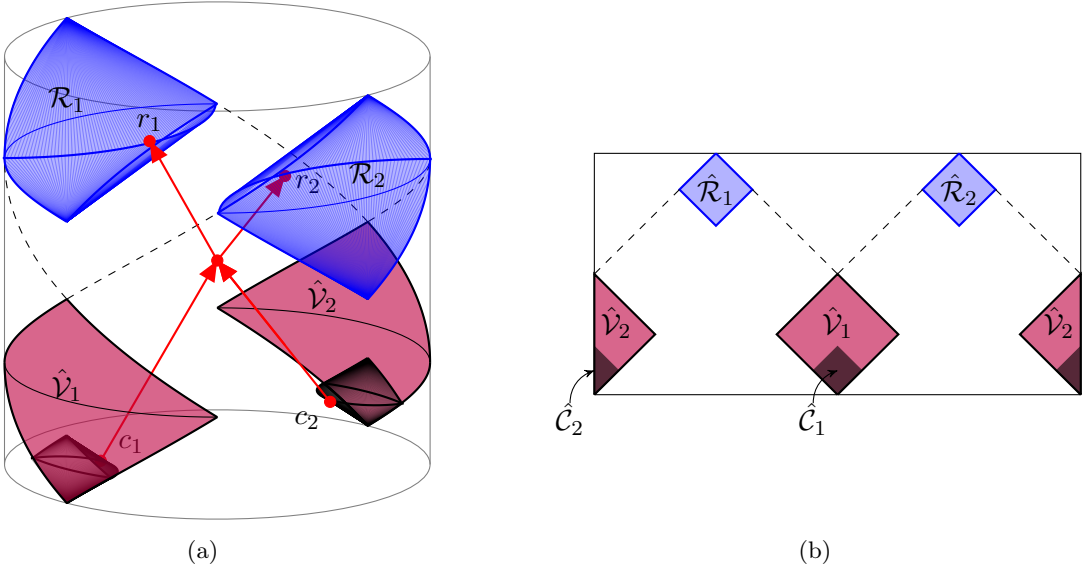

Relativistic quantum tasks are quantum computations with inputs and outputs occurring at specified spacetime locations \cite{kent2012quantum}. In the context of the AdS/CFT correspondence, the framework of holographic quantum tasks \cite{may2019quantum} considers such computations from a bulk as well as boundary perspective. By comparing the two perspectives, it has been possible to find a constraint on boundary entanglement placed by bulk causal features \cite{may2019quantum,may2020holographic}. This constraint is called the connected wedge theorem; it states that a specific entanglement wedge must be connected when a related set of light cones overlap. In this paper we extend the holographic quantum tasks framework and derive a stronger connected wedge theorem.

To derive constraints for AdS/CFT from quantum tasks, we begin by defining a task in the bulk. This is specified by sets of input and output locations, along with a channel relating inputs to outputs. We can then discuss protocols for completing this bulk task, and determine the probability with which the best such protocol succeeds. Then, we identify a corresponding task in the boundary, and note that because the boundary describes the bulk the boundary task is completed with the same or higher probability. In some cases we can, beginning with the success probability, show that there must be large amounts of entanglement in the boundary. Reasoning in this way leads to the connected wedge theorem. 

In \cite{may2019quantum,may2020holographic} the tasks in the bulk considered always had inputs and outputs taking a special form. First, the inputs and outputs were given at locations idealized as points, and second, those points were located at asymptotic infinity. This was convenient in that it allowed the corresponding boundary task to be identified with no additional assumptions: we can naturally identify a point in AdS at infinity with a point in the CFT, and in this way define the input and output locations for the boundary task. 

In this paper we extend this procedure to include bulk tasks with input locations that are still points, but which are not at asymptotic infinity. In the boundary, entanglement wedge reconstruction implies the corresponding task is one with extended regions as input locations, namely regions which have the bulk input point in their entanglement wedge. A similar relationship relates bulk output points and boundary output regions. An example set-up is shown in figure \ref{fig:introfig}.

By considering this broader class of holographic quantum tasks we are led to a stronger version of the connected wedge theorem than was previously given. A simple adaptation of the relativistic proof given for the earlier connected wedge theorem in \cite{may2020holographic} can be used to prove our theorem. As well, the new connected wedge theorem applies non-trivially to Poincar\'e-AdS$_{2+1}$, where the earlier theorem was always trivial. The application to Poincar\'e requires considering output regions which consist of two intervals, and so in particular are disconnected. This gives a particularly clear example of how the restriction of input and output locations to be points failed to capture all useful constructions. 

Note that this article emphasizes the quantum tasks perspective, however, the reader interested in the connected wedge theorem as a geometric statement can move directly to the theorem statement in section \ref{sec:newtheorem} and its relativistic proof in section \ref{sec:GRperspectiveonCWtheorem}. 

Throughout the article we refer to boundary spacetime regions with hatted script letters $\hat{\mathcal{X}},\hat{\mathcal{A}},\hat{\mathcal{B}},\hat{\mathcal{V}}...$ and their bulk entanglement wedges with un-hatted letters $\mathcal{E}_{W}(\hat{\mathcal{X}})=\mathcal{X}$, etc. We will use $J^\pm(\cdot)$ for the causal future or past taken in the bulk geometry, and $\hat{J}^\pm(\cdot)$ for the causal future or past taken in the boundary geometry. We will denote the Ryu-Takayanagi surface of a boundary region $\hat{\mathcal{X}}$ by $\gamma_{\mathcal{X}}$.

\subsection*{Outline of the article}

The outline of this article is as follows.

In section \ref{sec:framework}, we update the quantum tasks framework to consider input and output spacetime regions. We discuss in detail how to identify bulk and boundary tasks in this context. We emphasize that doing so requires identifying, for a given boundary region, a bulk region which stores the same quantum information. In other work this has been understood to be the entanglement wedge \cite{CKNR, HHLR, maximin, JLMS, DHW, noisyDHW}.

In section \ref{sec:newtheorem} we state the improved connected wedge theorem, making use of input and output regions. We explain why this is a stronger theorem than given previously, and comment on how to choose the inputs and output regions to arrive at the strongest possible statement. We also point out that the converse to the theorem does not hold. Finally we explain how to apply the improved connected wedge theorem to Poincar\'e-AdS$_{2+1}$, which involves taking one of the output regions to be disconnected.

In section \ref{sec:QTonCWtheorem} we give the quantum tasks argument for the improved connected wedge theorem, which exploits the expanded quantum tasks framework developed in section \ref{sec:framework}. Aside from the generalization to input and output regions, the treatment here also improves on \cite{may2020holographic} in the way errors in the bulk protocol are handled, in particular we derive a linear lower bound on the mutual information even in this noisy case. 

In section \ref{sec:GRperspectiveonCWtheorem} we prove the stronger connected wedge theorem using the focusing theorem in general relativity. The relativistic proof is a simple modification of the proof of the earlier theorem appearing in \cite{may2020holographic}.

We conclude with a brief summary and some comments in section \ref{sec:discussion}. 

\section{The holographic quantum tasks framework}\label{sec:framework}

\subsection{Quantum tasks}

We will discuss quantum tasks where Alice is given inputs that are initially recorded into extended spacetime regions, and must be output at extended output regions. To make this more precise, we define a notion of quantum information being localized to a spacetime region. Our definitions are adapted from \cite{hayden2019localizing}. 
\begin{definition}\label{def:localized}
Suppose one party, Alice, holds system $X$ of a quantum state $\ket{\Psi}_{XX'}$. Then we say the subsystem $X$ is \textbf{localized} to a spacetime region $\mathcal{R}$ if a second party, Bob, for whom the state is initially unknown can prepare the $X$ system by acting on $\mathcal{R}$ with some channel $\mathcal{M}_{\mathcal{R}\rightarrow X}$. 
\end{definition}
If system $X$ is localized to region $\mathcal{R}$ such that a channel $\mathcal{M}_{\mathcal{R}\rightarrow X}$ recovers $X$, we will say that $X$ is \textbf{localized to $\mathcal{R}$ relative to $\mathcal{M}_{\mathcal{R}\rightarrow X}$.} Note that throughout this article by quantum system we mean a tensor factor of a Hilbert space, e.g. the $X$ system of $\mathcal{H}_{XX'}=\mathcal{H}_X\otimes \mathcal{H}_{X'}$.\footnote{It may also be interesting to repeat our discussion in the more general setting where $X$ is a subalgebra.}

It will also be convenient to say a quantum system $X$ is \textbf{excluded} from a spacetime region $\mathcal{R}$ if Bob cannot learn anything about $X$ by accessing $\mathcal{R}$. One way to specify this precisely is to consider $\ket{\Psi}_{XX'}$ to be in the maximally entangled state. Then $X$ is excluded from $\mathcal{R}$ when $I(\mathcal{R}:\mathcal{X'})=0$. 

We should point out several features of these definitions. First, note that a system $X$ is localized to $\mathcal{R}$ if and only if it is localized to the domain of dependence of $\mathcal{R}$, and similarly it is excluded from $\mathcal{R}$ if and only if it is excluded from the domain of dependence of $\mathcal{R}$. This follows because all the classical and quantum data in the domain of dependence of $\mathcal{R}$ is fixed by the data in $\mathcal{R}$ by time evolution. Consequently we will identify regions with their domains of dependence throughout this paper. 

Second, note that a quantum system $X$ can be neither localized to nor excluded from a region $\mathcal{R}$ if some but not all information about $X$ is available in $\mathcal{R}$. As well, notice that given a Cauchy surface $\Sigma$ a quantum system can be excluded from both $\Sigma \cap \mathcal{R}$ and $\Sigma\setminus \mathcal{R}$. To do this, encode system $X$ using the one-time pad \cite{ambainis2000private} using a classical key $k$. This hides the state on $X$, which can only be revealed if $k$ is known. The $X$ register can then be passed through $\Sigma \cap \mathcal{R}$ and $k$ through $\Sigma \setminus \mathcal{R}$, and system $X$ will be excluded from both regions. Of course, $X$ will still be localized to the full Cauchy surface $\Sigma$. 

There are many possible ways in which a quantum system can be localized to a given spacetime region, and a single quantum system can be localized to many different spacetime regions. Which sets of spacetime regions the same quantum information can be localized to is restricted however. In particular, if $X$ is localized to a region $\mathcal{R}$ then it follows that $X$ is excluded from the spacelike complement $\mathcal{R}^c$. This is because otherwise we could act independently on $\mathcal{R}$ and $\mathcal{R}^c$ to produce copies of $X$, in violation of the no-cloning theorem. 

As a simple example of how a single quantum system can be localized to many regions, suppose we have three subregions $\mathcal{X}_1,\mathcal{X}_2,\mathcal{X}_3$ which are all spacelike separated. Then define $\mathcal{R}_1 = \mathcal{X}_1\cup \mathcal{X}_2$, $\mathcal{R}_2= \mathcal{X}_2\cup \mathcal{X}_3$, and $\mathcal{R}_3=\mathcal{X}_3\cup \mathcal{X}_1$. To localize a quantum system $X$ to all three regions $\{\mathcal{R}_1,\mathcal{R}_2,\mathcal{R}_3\}$, encode $X$ into an error-correcting code on three subsystems $S_1S_2S_3$ that corrects one erasure error. Send system $S_i$ to $\mathcal{X}_i$. Then each of the $\mathcal{R}_i$ contain two of the $S_i$ subsystems, and $X$ can be recovered from each of them. 

Given a quantum system $X$, we can specify how it is localized in spacetime by specifying two sets of regions, call them $\{\mathcal{A}_{X}^i\}_i$ and $\{\mathcal{U}_{X}^i\}_i$. We specify that $X$ is localized to each of the regions $\mathcal{A}_{X}^i$, and excluded from each of the regions $\mathcal{U}_X^i$. Implicitly, each region $\mathcal{A}_{X}^i$ comes with a channel $\mathcal{N}_{\mathcal{A}_{X}^i\rightarrow X}$ that specifies how $X$ can be recovered from $\mathcal{A}^i_X$.\footnote{In general there could be many such channels, though this will not be important in this article.} We summarize this in the next definition. 
\begin{definition}
A quantum system $X$ is encoded into an \textbf{access structure} $\mathcal{S}_X= (\{\mathcal{A}_X^i\}_i,\{\mathcal{U}_X^i\}_i)$ if $X$ is localized to each of the regions $\{\mathcal{A}_X^i\}_i$ and excluded from each of the regions $\{\mathcal{U}_X^i\}_i$.
\end{definition}
The term ``access structure'' is borrowed from the subject of quantum secret sharing, which features a closely related object.\footnote{In quantum secret sharing a system $X$ is recorded into shares $X_1...X_n$ such that $X$ can be recovered from some subsets of shares, called authorized sets, and no information about $X$ is available in another set of subsets of shares, called the unauthorized sets.} In \cite{hayden2019localizing} the authors characterized which access structures it is possible to localize a quantum system to. 

Next, we give a definition of a relativistic quantum task. This builds on the definition presented in \cite{may2019quantum} by allowing for inputs and outputs to be recorded into arbitrary access structures. 
\begin{definition}
A \textbf{relativistic quantum task} is defined by a tuple $\mathbf{T}=(\mathcal{M},\mathscr{A},\mathcal{S}_{\mathscr{A}},\mathscr{B},\mathcal{S}_{\mathscr{B}}, \mathcal{N}_{\mathscr{A}\rightarrow \mathscr{B}}
)$, where:
\begin{itemize}
    \item $\mathcal{M}$ is the spacetime in which the task occurs, it is described by a manifold equipped with a (Lorentzian) metric.
    \item $\mathscr{A}=A_1...A_{n_a}$ is the collection of all the input quantum systems, and $\mathcal{S}_{\mathscr{A}}=\{ \mathcal{S}_{A_1},...,\mathcal{S}_{A_{n_a}}\}$ is the set of all access structures for the input systems. 
    \item $\mathscr{B}=B_1...B_{n_b}$ is the collection of all the output quantum systems, and $\mathcal{S}_{\mathscr{B}}=\{ \mathcal{S}_{B_1},...,\mathcal{S}_{B_{n_b}}\}$ is the set of all access structures for the output systems. 
    \item $\mathcal{N}_{\mathscr{A}\rightarrow \mathscr{B}}$ is a quantum channel that maps the input systems $\mathscr{A}$ to the output systems $\mathscr{B}$.
\end{itemize}
Bob encodes the input systems $A_i$ in such a way that the access structures $\mathcal{S}_{A_i}$ are satisfied. To complete the task, Alice should apply the channel $\mathcal{N}_{\mathscr{A}\rightarrow \mathscr{B}}$ and localize each of the systems $B_i$ according to the access structure $\mathcal{S}_{{B}_i}$. 
\end{definition}
In order to encode the $A_i$ into the appropriate regions, Bob couples the regions $\mathcal{A}_{A_i}^j$ to some external system which initially hold the $A_i$. To verify Alice has completed the task successfully, Bob will access one or more of the regions $\mathcal{A}_{B_j}^i$, $\mathcal{U}_{B_j}^i$ and attempt to recover system $B_j$. If Bob is able to produce $B_j$ from the authorized region $\mathcal{A}_{B_j}^i$ he declares the task successful. Similarly if he is \emph{unable} to produce $B_j$ from the unauthorized region $\mathcal{U}_{B_j}^i$ he declares the task successful. The probability that Alice's outputs pass Bob's test is her success probability. Alice's success probability maximized over all possible protocols for completing the task is the tasks success probability, $p_{suc}(\mathbf{T})$. 

Note that if Bob acts on one of the output regions $\mathcal{A}_{B_j}^i$, $\mathcal{U}_{B_j}^i$ in performing his test, we no longer require Alice have the correct outputs (or exclusions from) regions in the causal future of the accessed region. Similarly, Bob will localize the inputs $A_j$ to regions $\mathcal{A}_{A_j}^i$ so long as Alice never interferes. She may choose to access some region $\mathcal{A}_{A_j}^i$ however and obtain $A_i$, in which case Bob is no longer expected to localize $A_i$ to regions in the future of $\mathcal{A}_{A_j}^i$.

In the application considered below, we begin with a spacetime and use tasks as a way to probe features of that fixed geometry. Consequently, we have defined quantum tasks to feature a fixed spacetime background. Doing so assumes Alice's choice of protocol does not change the geometry. It is also possible to consider more general tasks, where we allow the spacetime geometry to react to Alice's protocol, which might for instance involve distributing large numbers of qubits which then change the geometry. We leave considering this to future work. 

\subsection{Quantum tasks in holography}\label{subsec:holographictasks}

In our definition of a quantum task in the last section, we have used an operational framing. This is only for convenience however, and it is possible to remove this language. In particular, the protocol Alice carries out is in fact just a feature of some initial state $\ket{\Psi}$. All the instructions for her protocol are by necessity recorded there, all that happens during the execution of the protocol is time evolution according to the underlying theory's Hamiltonian. While Alice's protocol is the internal dynamics of the theory in question, Bob preparing the inputs and collecting outputs correspond to couplings to some external system. Viewing quantum tasks in this way motivates understanding them as probes of the underlying theory they are defined in. 

Because tasks probe the underlying theory, if we are given an equivalence between two theories it is natural to try and interpret this equivalence in the language of tasks. In particular we will consider the bulk and boundary theories in AdS/CFT. Within each theory, there is a set of tasks that can be defined and associated success probabilities, $\{(\mathbf{T}_i,p_{suc}(\mathbf{T}_i))_i\}$. The equivalence of bulk and boundary theories suggests that for each task $\mathbf{T}$ defined in the bulk there is some corresponding task $\mathbf{\hat{T}}$ in the boundary, and further that $p_{suc}(\mathbf{\hat{T}}) = p_{suc}(\mathbf{{T}})$. We will make this more precise below.

Our first step will be to restrict attention to a bulk described by classical geometry along with quantum fields living on a curved background (which may be coupled to the geometry). This means that while the boundary theory completely describes the bulk, the converse is not true. Consequently we will expect an inequality, $p_{suc}(\mathbf{{T}}) \leq p_{suc}(\mathbf{\hat{T}})$. Before understanding this in more detail however, we need to specify how a bulk task should be associated with a boundary task. 

Given a task in the bulk $\mathbf{T}=(\mathcal{M},\mathscr{A},\mathcal{S}_{\mathscr{A}},\mathscr{B},\mathcal{S}_{\mathscr{B}}, \mathcal{N}_{\mathscr{A}\rightarrow \mathscr{B}})$, we should identify the boundary dual of each element of the tuple. Beginning with $\mathcal{M}$, the bulk geometry, we define $\mathbf{\hat{T}}$ to be in the geometry $\partial \mathcal{M}$, the boundary of $\mathcal{M}$. The inputs $\mathscr{A}$, outputs $\mathscr{B}$, and channel $\mathcal{N}_{\mathscr{A}\rightarrow \mathscr{B}}$ we may identify trivially across bulk and boundary. This is because while the bulk and boundary degrees of freedom look very different, we can record the same quantum states into these different degrees of freedom.

Next we need to discuss how to identify an access structure in the bulk with a corresponding access structure in the boundary. Note that in principle, because the AdS/CFT correspondence fixes the boundary description given the bulk, the boundary access structure $(\{\hat{\mathcal{A}}_{A_i}^j\},\{\hat{\mathcal{U}}_{A_i}^j\}\}\})$ is fixed by the bulk one $(\{{\mathcal{A}}_{A_i}^j\},\{{\mathcal{U}}_{A_i}^j\}\}\})$. We have not understood how to do this in the most general case, but can make some statements which will be sufficient for the application discussed here. 

First, notice that given bulk authorized regions $\{\mathcal{A}_{A_i}^j\}_j$, we have
\begin{align}\label{eq:authorizedinclusion}
    \{\hat{\mathcal{A}}_{A_i}^k\}_k \supseteq \bigcup_{j} \{ \hat{X}: {\mathcal{A}}_{A_i}^j\subseteq \mathcal{E}_W(\hat{X})\}.
\end{align}
This is because the entanglement wedge $\mathcal{E}_W(\hat{X})$ is the portion of the bulk which $\hat{X}$ can be used to recover \cite{CKNR, HHLR, maximin, JLMS, DHW, noisyDHW}, so when ${\mathcal{A}}_{A_i}^j\subseteq \mathcal{E}_W(\hat{X})$ the boundary region $\hat{X}$ can be used to recover $A_i$, which implies $\hat{X}$ is an authorized region. The other inclusion does not follow in general since there may be some boundary regions $\hat{\mathcal{A}}_{A_i}^j$ whose entanglement wedge includes a portion but not all of $\mathcal{A}_{A_i}^j$ and which still construct $A_i$. 

Given a bulk unauthorized region, we can say that
\begin{align}
    \{\hat{\mathcal{U}}_{A_i}^k\}_k \supseteq \bigcup_j \{ \hat{X}: {\mathcal{U}}_{A_i}^j\supseteq \mathcal{E}_W(\hat{X})\}.
\end{align}
This follows because $\mathcal{E}_W(\hat{X})$ is the largest bulk region whose quantum information can be reconstructed given $\hat{X}$, so ${\mathcal{U}}_{A_i}^j\supseteq \mathcal{E}_W(\hat{X})$ means $\hat{X}$ does not reconstruct $A_i$. Note that unless one or more of the ${\mathcal{U}}_{A_i}^j$ are anchored to the boundary, the set $\{ \hat{X}: {\mathcal{U}}_{A_i}^j\supseteq \mathcal{E}_W(\hat{X})\}$ will be empty. 

We will be interested only in a special case, where the bulk tasks access structures all have only authorized regions, and where those authorized regions are points. In this case the inclusion \ref{eq:authorizedinclusion} becomes an equality, fully specifying the boundary authorized regions from the bulk ones. Further, there will be no boundary unauthorized regions.\footnote{Of course the spacelike complements $[\mathcal{A}_{A_i}^k]^c$ do not contain any information about $A_i$, but it is not necessary to designate these as unauthorized, since this is immediate from the $\mathcal{A}_{A_i}^k$ being authorized.}

Given a bulk task $\mathbf{T}$ and associated boundary task $\mathbf{\hat{T}}$, we've claimed $p_{suc}(\mathbf{{T}}) \leq p_{suc}(\mathbf{\hat{T}})$. This follows because any protocol that completes the task in the bulk with some probability $p$ will be mapped under the AdS/CFT duality to a protocol in the boundary. The bulk task's success probability is determined by the information localized to the regions $(\{{\mathcal{A}}_{A_i}^j\},\{{\mathcal{U}}_{A_i}^j\}\}\})$. In the boundary description the same information will be available in the regions $(\{\hat{\mathcal{A}}_{A_i}^j\},\{\hat{\mathcal{U}}_{A_i}^j\}\}\})$, so the boundary protocol will complete the task with probability $p$ as well. Note that we claim only an inequality, rather than an equality, because many protocols in the boundary theory will correspond to bulk protocols that change the geometry $\mathcal{M}$, which we assumed should be fixed and unaffected by the protocol. Worse, some boundary protocols might correspond to leaving a semi-classical description of the bulk altogether. 

\section{An improved connected wedge theorem}\label{sec:newtheorem}

\subsection{Statement of the theorem}

\begin{figure}
    \centering
    \subfloat[\label{fig:boundaryregions}]{
    \begin{tikzpicture}[scale=1.6]

    \draw (-2,0) -- (2,0) -- (2,2) -- (-2,2) -- (-2,0);
    
    \draw[fill=black!60!,opacity=0.8] (0,0) -- (0.5,0.5) -- (0,1) -- (-0.5,0.5) -- (0,0);
    \draw[thick] (0,0) -- (0.5,0.5) -- (0,1) -- (-0.5,0.5) -- (0,0);
    \node at (0,0.5) {$\hat{\mathcal{C}}_1$};
    
    \draw[fill=black!60!,opacity=0.8] (-2,0) -- (-1.5,0.5) -- (-2,1) -- (-2,0);
    \draw[thick] (-2,0) -- (-1.5,0.5) -- (-2,1) -- (-2,0);
    \draw[fill=black!60!,opacity=0.8] (2,0) -- (1.5,0.5) -- (2,1) -- (2,0);
    \draw[thick] (2,0) -- (1.5,0.5) -- (2,1) -- (2,0);
    
    \node at (1.82,0.5) {$\hat{\mathcal{C}}_2$};
    \node at (-1.82,0.5) {$\hat{\mathcal{C}}_2$};
    
    \draw[fill=blue,opacity=0.3] (-1,2) -- (-2,1) -- (-1,0) -- (0,1) -- (-1,2);
    \draw[thick,blue] (-1,2) -- (-2,1) -- (-1,0) -- (0,1) -- (-1,2);
    
    \node at (1,1) {$\hat{\mathcal{R}}_2$};
    
    \draw[fill=blue,opacity=0.3] (1,2) -- (2,1) -- (1,0) -- (0,1) -- (1,2);
    \draw[thick,blue] (1,2) -- (2,1) -- (1,0) -- (0,1) -- (1,2);
    
    \node at (-1,1) {$\hat{\mathcal{R}}_1$};
    
    \node at (0,-1) {$ $};
    
    \end{tikzpicture}
    }
    \hfill
    \centering
    \subfloat[\label{fig:connected-scattering}]{
    \tdplotsetmaincoords{15}{0}
    \begin{tikzpicture}[scale=1.4,tdplot_main_coords]
    \tdplotsetrotatedcoords{0}{45}{0}
    \draw[gray] (-2,1,0) -- (-2,5,0);
    \draw[gray] (2,1,0) -- (2,5,0);
    
    \begin{scope}[tdplot_rotated_coords]
    
    \draw[domain=0:45,variable=\x,smooth, fill=black!60!,opacity=0.8] plot ({-2*sin(\x)}, {1+\x/45}, {2*cos(\x)}) -- plot ({-2*sin((45-\x))}, {3-(45-\x)/45}, {2*cos(45-\x)}) --  plot ({2*sin(\x)}, {3-\x/45}, {2*cos(\x)}) -- plot ({2*sin(45-\x)}, {1+(45-\x)/45}, {2*cos(45-\x)});
    
    \draw[domain=0:45,variable=\x,smooth,thick] plot ({-2*sin(\x)}, {1+\x/45}, {2*cos(\x)});
    \draw[domain=0:45,variable=\x,smooth,thick] plot ({2*sin(\x)}, {1+\x/45}, {2*cos(\x)});
    \draw[domain=0:45,variable=\x,smooth,thick] plot ({-2*sin(\x)}, {3-\x/45}, {2*cos(\x)});
    \draw[domain=0:45,variable=\x,smooth,thick] plot ({2*sin(\x)}, {3-\x/45}, {2*cos(\x)});
    
    \begin{scope}[canvas is xz plane at y=1]
    \draw[gray] (0,0) circle[radius=2] ;
    \end{scope}
    
    \begin{scope}[canvas is xz plane at y=5]
    \draw[gray] (0,0) circle[radius=2] ;
    \end{scope}
    
    \begin{scope}[canvas is xz plane at y=2]
    \draw[gray] (0,0) circle (2);
    \end{scope}
    
    \draw [domain=-45:45] plot ({2*cos(\x+90)},2, {2*sin(\x+90)});
    
    \draw[domain=0:90,variable=\x,smooth,blue] plot ({2*sin(\x+180)}, {3+\x/45}, {2*cos(\x+180)});
    \draw[domain=0:90,variable=\x,smooth,blue] plot ({2*sin(\x+180)}, {(3-\x/45)}, {2*cos(\x+180)});
    
    \draw[domain=0:90,variable=\x,smooth,blue] plot ({2*sin(\x+180)}, {3+\x/45}, {-2*cos(\x+180)});
    \draw[domain=0:90,variable=\x,smooth,blue] plot ({2*sin(\x+180)}, {(3-\x/45)}, {-2*cos(\x+180)});
    
    \draw[blue] ({2*sin(180)}, {3}, {-2*cos(180)}) -- ({2*sin(180)}, {3}, {2*cos(180)});
    
    \foreach \x in {0,...,90}
    {
    \draw[blue,smooth,opacity=0.3] ({2*sin(\x+180)}, {3+\x/45}, {2*cos(\x+180)})-- ({2*sin(\x+180)}, {3-\x/45}, {2*cos(\x+180)});
    }
    
    \foreach \x in {0,...,89}
    {
    \draw[blue,smooth,opacity=0.3] ({2*sin(\x+180)}, {3+\x/45}, {-2*cos(\x+180)}) -- ({2*sin(\x+180)}, {3-\x/45}, {-2*cos(\x+180)});
    }
    
    \foreach \x in {0,...,90}
    {
    \draw[blue,smooth,opacity=0.5] ({2*sin(\x+180)}, {3+\x/45}, {2*cos(\x+180)}) -- ({2*sin(\x+180)}, {3+\x/45}, {-2*cos(\x+180)});
    }
    
    \foreach \x in {0,...,90}
    {
    \draw[blue,smooth,opacity=0.3] ({2*sin(\x+180)}, {3-\x/45}, {2*cos(\x+180)}) -- ({2*sin(\x+180)}, {3-\x/45}, {-2*cos(\x+180)});
    }
    
    \draw[domain=0:90,variable=\x,smooth,dashed,blue] plot ({-2*sin(\x+180)}, {3+\x/45}, {2*cos(\x+180)});
    \draw[domain=0:90,variable=\x,smooth,dashed,blue] plot ({-2*sin(\x+180)}, {3+\x/45}, {-2*cos(\x+180)});
    \draw[domain=0:90,variable=\x,smooth,dashed,blue] plot ({-2*sin(\x+180)}, {3-\x/45}, {2*cos(\x+180)});
    \draw[domain=0:90,variable=\x,smooth,dashed,blue] plot ({-2*sin(\x+180)}, {3-\x/45}, {-2*cos(\x+180)});
    
    \draw[thick,red,-triangle 45] (0,2,-0.9) -- (0,3,0);
    \draw[red] plot [mark=*, mark size=1.5] coordinates{(0,2,-0.9)};
    
    \draw[thick,red,-triangle 45] (0,2,0.9) -- (0,3,0);
    
    \draw[domain=0:45,variable=\x,smooth, fill=black!50!,opacity=0.8] plot ({-2*sin(\x+180)}, {1+\x/45}, {2*cos(\x+180)}) -- plot ({-2*sin((45-\x)+180)}, {3-(45-\x)/45}, {2*cos(45-\x+180)}) --  plot ({2*sin(\x+180)}, {3-\x/45}, {2*cos(\x+180)}) -- plot ({2*sin(45-\x+180)}, {1+(45-\x)/45}, {2*cos(45-\x+180)});
    
    \begin{scope}[canvas is xz plane at y=2]
    \draw [domain=-180:180] plot ({2*cos(\x-90)}, {2*sin(\x-90)});
    \end{scope}
    
    \draw[ domain=0:45,variable=\x,smooth,thick] plot ({-2*sin(\x+180)}, {1+\x/45}, {2*cos(\x+180)});
    
    \draw[domain=0:45,variable=\x,smooth,thick] plot ({2*sin(\x+180)}, {1+\x/45}, {2*cos(\x+180)});
    
    \draw[domain=0:45,variable=\x,smooth,thick] plot ({-2*sin(\x+180)}, {3-\x/45}, {2*cos(\x+180)});
    
    \draw[domain=0:45,variable=\x,smooth,thick] plot ({2*sin(\x+180)}, {3-\x/45}, {2*cos(\x+180)});
    
    \draw[domain=45:135, thick] plot
    ({2*cos(\x)},{2},{2*sin(\x)-2.85});
    \draw[domain=45:135, thick] plot
    ({2*cos(\x)},{2},{-2*sin(\x)+2.85});
    
    \foreach \x in {45,...,135}
    {
    \draw[opacity=0.3] (0,3,2) -- ({2*cos(\x)},{2},{-2*sin(\x)+2.85});
    \draw[opacity=0.3] (0,3,-2) -- ({2*cos(\x)},{2},{2*sin(\x)-2.85});
    }
    
    \foreach \x in {45,...,135}
    {
    \draw[opacity=0.3] (0,1,2) -- ({2*cos(\x)},{2},{-2*sin(\x)+2.85});
    \draw[opacity=0.3] (0,1,-2) -- ({2*cos(\x)},{2},{2*sin(\x)-2.85});
    }
    
    \foreach \x in {11,14,...,41}
    {
    \draw[opacity=0.3] plot (0,1,-2) -- ({2*sin(\x+180)}, {1+(\x)/45}, {2*cos(\x+180)});
    }

    \foreach \x in {11,14,...,41}
    {
    \draw[opacity=0.3] plot (0,1,2) -- ({2*sin(\x+180)}, {1+(\x)/45}, {-2*cos(\x+180)});
    }
    
    \foreach \x in {11,14,...,41}
    {
    \draw[opacity=0.3] plot (0,1,2) -- ({-2*sin(\x+180)}, {1+(\x)/45}, {-2*cos(\x+180)});
    }
    
    \foreach \x in {11,14,...,41}
    {
    \draw[opacity=0.3] plot (0,3,2) -- ({-2*sin(\x+180)}, {3-(\x)/45}, {-2*cos(\x+180)});
    }
    
    \foreach \x in {11,14,...,41}
    {
    \draw[opacity=0.3] plot (0,3,-2) -- ({2*sin(\x+180)}, {3-(\x)/45}, {2*cos(\x+180)});
    }
    
    \foreach \x in {11,14,...,41}
    {
    \draw[opacity=0.3] plot (0,3,-2) -- ({-2*sin(\x+180)}, {3-(\x)/45}, {2*cos(\x+180)});
    }
    
    \draw[red] plot [mark=*, mark size=1.5] coordinates{(0,2,0.9)};
    \draw[red] plot [mark=*, mark size=1.5] coordinates{(0,3,0)};
    
    \node at (-2,1.4,0) {$\mathcal{C}_1$};
    \node at (-2,1.5,4) {$\mathcal{C}_2$};
    
    \node at (-1.5,3.5,0) {$\mathcal{R}_1$};
    
    \end{scope}
    \end{tikzpicture}
    }
    \caption{(a) A view of the boundary of AdS$_{2+1}$. Left and right edges of the diagram are identified. Shown is an example choice of input regions $\hat{\mathcal{C}}_i$ and output regions $\hat{\mathcal{R}}_i$. This particular choice is maximal for the regions $\hat{\mathcal{V}}_i$ defined by $\hat{\mathcal{C}}_i=\hat{\mathcal{V}}_i$. (b) Bulk perspective on the same choice of regions, showing the entanglement wedges $\mathcal{C}_i$ and $\mathcal{R}_i$. Only one of the out regions is shown, to avoid cluttering the diagram. In the bulk there is a non-empty entanglement scattering region $J^{\mathcal{E}}_{12\rightarrow 12}$.}
\end{figure}
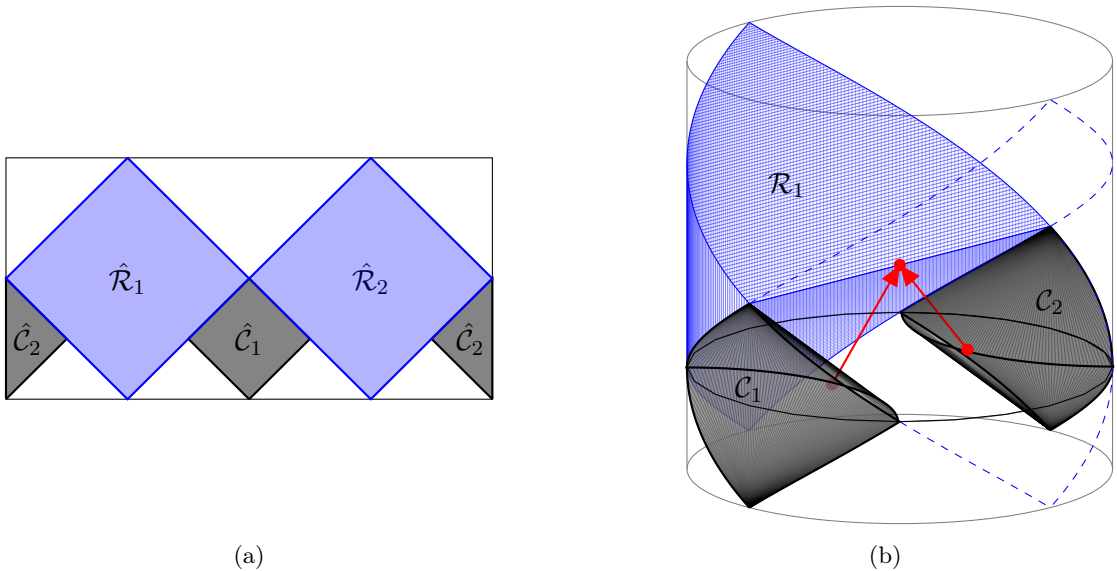

We state the improved connected wedge theorem below. 
\begin{theorem}\label{thm:main} \textbf{(Connected wedge theorem)} Pick four regions $\hat{\mathcal{C}}_1,\hat{\mathcal{C}}_2,\hat{\mathcal{R}}_1,\hat{\mathcal{R}}_2$ on the boundary of an asymptotically AdS spacetime. From these, define the decision regions
\begin{align}
    \hat{\mathcal{V}}_1 &\equiv \hat{J}^+(\hat{\mathcal{C}}_1) \cap \hat{J}^-(\hat{\mathcal{R}}_1) \cap \hat{J}^-(\hat{\mathcal{R}}_2) , \nonumber \\
    \hat{\mathcal{V}}_2 &\equiv \hat{J}^+(\hat{\mathcal{C}}_2) \cap \hat{J}^-(\hat{\mathcal{R}}_1) \cap \hat{J}^-(\hat{\mathcal{R}}_2).
\end{align}
Assume that $\hat{\mathcal{C}}_i \subseteq \hat{\mathcal{V}}_i$. Define the entanglement scattering region
\begin{align}
    J_{12\rightarrow 12}^{\mathcal{E}} \equiv J^+(\mathcal{C}_1)\cap J^+(\mathcal{C}_2) \cap J^-(\mathcal{R}_1) \cap J^-(\mathcal{R}_2),
\end{align}
where $\mathcal{C}_i = \mathcal{E}_W(\hat{\mathcal{C}}_i)$ and $\mathcal{R}_i = \mathcal{E}_W(\hat{\mathcal{R}}_i)$. Then, $J_{12\rightarrow 12}^{\mathcal{E}}\neq \varnothing$ implies that $\mathcal{E}_W(\hat{\mathcal{V}}_1\cup \hat{\mathcal{V}}_2)$ is connected. 
\end{theorem}
Notice that this theorem generalizes the earlier one appearing in \cite{may2020holographic}. In particular choosing the $\hat{\mathcal{C}}_i$ and $\hat{\mathcal{R}}_i$ to be points we recover the earlier theorem. Also note that the theorem is true in arbitrary dimensions, but is trivial whenever $\hat{\mathcal{V}}_1$ and $\hat{\mathcal{V}}_2$ overlap. We summarize what is known about where non-trivial configurations occur in section \ref{sec:non-trivial}. 

It is interesting to consider starting with a choice of regions $\hat{\mathcal{V}}_1$ and $\hat{\mathcal{V}}_2$, then pick regions $\hat{\mathcal{C}}_1,\hat{\mathcal{C}}_2,\hat{\mathcal{R}}_1,\hat{\mathcal{R}}_2$ to understand if $\hat{\mathcal{V}}_1$ and $\hat{\mathcal{V}}_2$ share a connected entanglement wedge. In AdS$_{2+1}$, and when the decision regions each consist of a single diamond, there is a unique `best' way to do this, in the sense that one particular choice of regions will conclude there is a connected wedge whenever any choice of regions does. 

To find the optimal choice of $\hat{\mathcal{C}}_i,\hat{\mathcal{R}}_i$, we note first that there is a maximal choice of regions $\hat{\mathcal{C}}_i$ imposed by the constraint $\hat{\mathcal{C}}_i\subseteq \hat{\mathcal{V}}_i$: choose $\hat{\mathcal{C}}_i=\hat{\mathcal{V}}_i$. Further, there is a maximal choice of $\hat{\mathcal{R}}_i$ consistent with a given $\hat{\mathcal{V}}_1,\hat{\mathcal{V}}_2$, which is illustrated in figure \ref{fig:boundaryregions}. Since any other choice $\hat{\mathcal{C}}_i',\hat{\mathcal{R}}_i'$ has $\hat{\mathcal{C}}_i'\subseteq \hat{\mathcal{C}}_i$ and $\hat{\mathcal{R}}_i'\subseteq \hat{\mathcal{R}}_i$ these maximal choices have ${J'}^{\mathcal{E}}_{12\rightarrow 12}\subseteq J^{\mathcal{\mathcal{E}}}_{12\rightarrow 12}$, so whenever a non-maximal choice has a non-empty entanglement scattering region the maximal choice will. Thus whenever $\hat{\mathcal{C}}_i',\hat{\mathcal{R}}_i'$ can be used to conclude $\hat{\mathcal{V}}_1\cup \hat{\mathcal{V}}_2$ has a connected entanglement wedge, the maximal choice will conclude the same. 

\begin{figure}
    \centering
    \subfloat[\label{fig:withoutmatter}]{
    \begin{tikzpicture}[scale=0.9]
    
    \draw[thick] (0,0) circle (3);
    
    \draw[blue, thick] (2.12, 2.12) to [out=-135,in=135] (2.12, -2.12);
    \draw[blue, thick] (-2.12, 2.12) to [out=-45,in=45] (-2.12, -2.12);
    
    \draw[red, thick] (-2.12, 2.12) to [out=-45,in=-135] (2.12, 2.12);
    \draw[red, thick] (-2.12, -2.12) to [out=45,in=135] (2.12, -2.12);
    
    \node[right] at (3,0) {$\hat{\mathcal{V}}_2$};
    \node[left] at (-3,0) {$\hat{\mathcal{V}}_1$};
    
    \draw[<->,gray,domain=135:45] plot ({3.3*cos(\x)},{3.3*sin(\x)});
    \node[above] at (0,3.3) {$\pi/2$};
    
    \end{tikzpicture}
    }
    \hfill
    \subfloat[\label{fig:withmatter}]{
    \begin{tikzpicture}[scale=0.9]
    
    \node[right] at (3,0) {$\hat{\mathcal{V}}_2$};
    \node[left] at (-3,0) {$\hat{\mathcal{V}}_1$};
    
    \draw[thick] (0,0) circle (3);
    \draw[fill=black!60!,opacity=0.5] (0,0) circle (1);
    
    \draw[blue, thick] (2.12, 2.12) to [out=-135,in=90] (1.5, 0);
    \draw[blue, thick] (1.5, 0) to [out=-90,in=135] (2.12, -2.12);
    
    \draw[red, thick] (2.12, 2.12) to [out=-135,in=0] (0, 1.5);
    \draw[red, thick] (0, 1.5) to [out=-180,in=-45] (-2.12, 2.12);
    
    \draw[red, thick] (2.12, -2.12) to [out=135,in=0] (0, -1.5);
    \draw[red, thick] (0, -1.5) to [out=-180,in=45] (-2.12, -2.12);
    
    \draw[blue, thick] (-2.12, 2.12) to [out=-45,in=90] (-1.5, 0);
    \draw[blue, thick] (-1.5, 0) to [out=-90,in=45] (-2.12, -2.12);
    
    \end{tikzpicture}
    }
    \caption{A counterexample to the converse of Theorem \ref{thm:main}. (a) Vacuum AdS$_{2+1}$ with regions $\hat{\mathcal{V}}_1$ and $\hat{\mathcal{V}}_2$ chosen antipodally and to each occupy $\pi/2$ of the boundary. Choosing the maximal consistent input and output regions, the entanglement scattering region is exactly one point, and the Ryu-Takayanagi surface is on the transition from disconnected (blue) to connected (red). (b) Spherically symmetric matter is added to the bulk. Now the entanglement wedges of $\hat{\mathcal{V}}_1$ and $\hat{\mathcal{V}}_2$ reach less deeply into the bulk \cite{hubeny2012extremal}, and the light rays sent inward normally from their extremal surfaces are delayed. This closes the entanglement scattering region. By spherical symmetry however the Ryu-Takayanagi surface remains on the transition. Deforming $\hat{\mathcal{V}}_1$ to be larger ensures we are in the connected phase, and for small enough deformation ensures the scattering region remains empty. Figure reproduced from \cite{may2020holographic}.}
    \label{fig:counterexample}
\end{figure}
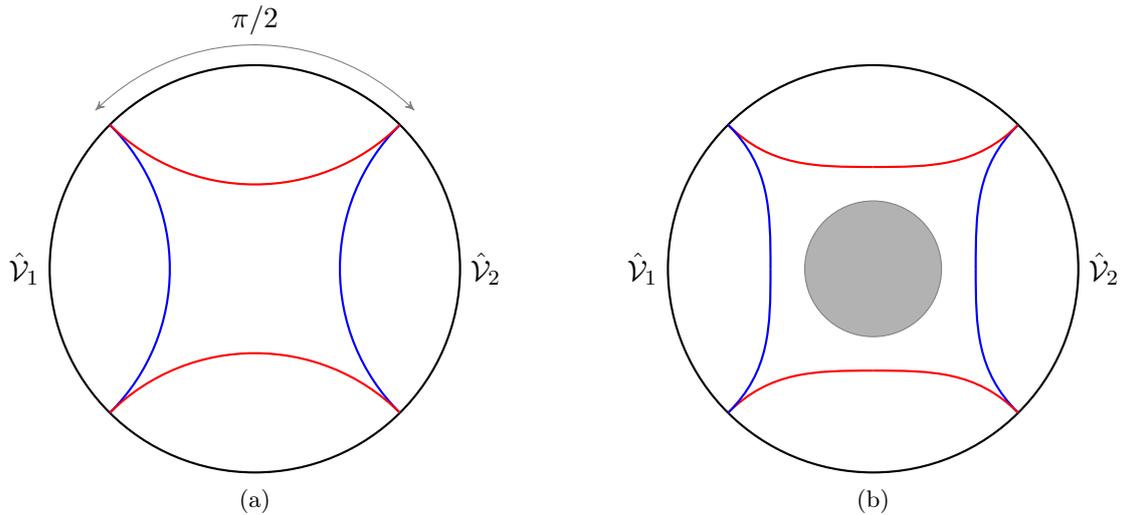

In general, the converse to Theorem \ref{thm:main} does not hold. This is immediate from the fact, pointed out in \cite{may2020holographic}, that the converse to the point-based case does not hold, which is a special case of the theorem presented here. As commented on in the last paragraph however, the point based choice is not the strongest choice of regions to understand if the entanglement wedge $\mathcal{E}_W(\hat{\mathcal{V}}_1\cup \hat{\mathcal{V}}_2)$ is connected. We can only expect a converse in the case where we take the optimal choice of regions outlined above. Taking the optimal choice of regions however the theorem still does not have a converse, as we argue in figure \ref{fig:counterexample}.

\subsection{Connected wedge theorem in Poincar\'e-AdS\texorpdfstring{$_{2+1}$}{TEXT}}\label{sec:planar}

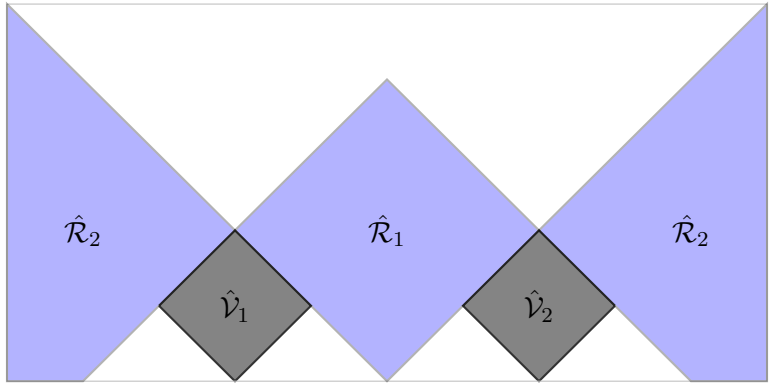
\begin{figure}
    \centering
    \begin{tikzpicture}[scale=0.5]
    
    \draw[thick, black, fill=black!60!,opacity=0.8] (-6,0) -- (-4,2) -- (-2,0) -- (-4,-2) -- (-6,0);
    
    \draw[thick, black, fill=black!60!,opacity=0.8] (6,0) -- (4,2) -- (2,0) -- (4,-2) -- (6,0);
    
    \draw[lightgray] (-10,8) -- (10,8) -- (10,-2) -- (-10,-2) -- (-10,8);

    \draw[thick,fill=blue,opacity=0.3] (-4,2) -- (0,6) -- (4,2) -- (0,-2) -- (-4,2);
    \draw[thick,fill=blue,opacity=0.3] (-4,2) -- (-10,8) -- (-10,-2) -- (-8,-2) -- (-4,2);
    \draw[thick,fill=blue,opacity=0.3] (4,2) -- (10,8) -- (10,-2) -- (8,-2) -- (4,2);
    
    \node at (4,0) {$\hat{\mathcal{V}}_2$};
    \node at (-4,0) {$\hat{\mathcal{V}}_1$};

    \node at (0,2) {$\hat{\mathcal{R}}_1$};
    \node at (-8,2) {$\hat{\mathcal{R}}_2$};
    \node at (8,2) {$\hat{\mathcal{R}}_2$};
    
    \end{tikzpicture}
    \caption{A typical choice of regions $\hat{\mathcal{C}}_1=\hat{\mathcal{V}}_1$, $\hat{\mathcal{C}}_2=\hat{\mathcal{V}}_2$, $\hat{\mathcal{R}}_1, \hat{\mathcal{R}}_2$ in the boundary of Poincar\'e-AdS which leads to a non-trivial conclusion in the connected wedge Theorem \ref{thm:main}. Region $\hat{\mathcal{R}}_2$ consists of two wedges which each extend to infinity.}
    \label{fig:poincaresetupmain}
\end{figure}

The connected wedge theorem applies to any asymptotically AdS spacetime, including global AdS and Poincar\'e-AdS spacetimes in arbitrary dimensions. To apply the theorem meaningfully however, we need to find configurations of regions $\hat{\mathcal{C}}_1,\hat{\mathcal{C}}_2,\hat{\mathcal{R}}_1,\hat{\mathcal{R}}_2$ such that the bulk entanglement scattering region is non-empty, while the boundary scattering region is empty.

It is not immediately clear how to find non-trivial configurations of input and output regions in Poincar\'e AdS$_{2+1}$. Indeed, at least for pure Poincar\'e-AdS$_{2+1}$ no such configurations exist when the input and output regions are chosen to be points. One way to see this is to start with non-trivial arrangements of points $c_1,c_2,r_1,r_2$ in global AdS$_{2+1}$, and chose a Poincar\'e patch which includes regions $\hat{\mathcal{V}}_1$ and $\hat{\mathcal{V}}_2$. Doing so one always finds that one of the four points sit outside the patch, and consequently we cannot state the non-trivial instances of the theorem using points directly in Poincar\'e AdS$_{2+1}$. 

For extended regions it is straightforward to find non-trivial configurations in Poincar\'e AdS$_{2+1}$. An example configuration is shown in figure \ref{fig:poincaresetupmain}. Importantly, region $\hat{\mathcal{R}}_2$ consists of two disconnected parts, where each connected component consists of a half line. In appendix \ref{sec:poincareexamples} we find configurations which are non-trivial in the case where the bulk is pure AdS. Since many of these configurations have extended entanglement scattering regions, and the scattering regions should be deformed only a small amount for small perturbations to the bulk geometry, there will also be many non-trivial configurations when matter is added. 

\section{Quantum tasks perspective on the connected wedge theorem}\label{sec:QTonCWtheorem}

\subsection{The \texorpdfstring{$\mathbf{{B}}_{84}^{\times n}$}{TEXT} task}\label{sec:B84task}

Following \cite{may2019quantum,may2020holographic}, we discuss the $\mathbf{B}_{84}$ task.\footnote{The name of this task comes from its similarity to the BB84 key distribution protocol, itself named for Bennet and Brassard \cite{bennett2020quantum}.} This task has $\mathcal{C}_1$ and $\mathcal{C}_2$ as authorized regions for inputs $A_1$ and $A_2$, and $\mathcal{R}_1$ and $\mathcal{R}_2$ as authorized regions for outputs $B_1$ and $B_2$. Alice will be given a guarantee that $A_1$ is in one of the states $H^q\ket{b}$, and $A_2$ stores the classical data $q$. Both $q$ and $b$ are bits, $q,b\in \{0,1\}$. Alice's task is to localize $b$ to both $\mathcal{R}_1$ and $\mathcal{R}_2$.

\begin{figure}
    \centering
    \subfloat[\label{fig:localschematic}]{
    \begin{tikzpicture}[scale=0.7]
    
    \draw[postaction={on each segment={mid arrow}}] (-4,0) -- (0,4) -- (-4,8);
    \draw[postaction={on each segment={mid arrow}}] (4,0) -- (0,4) -- (4,8);
    
    \draw[fill=yellow] (0,4) circle (0.3);
    
    \node[below] at (-4,-0.1) {$\mathcal{C}_1$};
    \draw[fill=black] (-4,0) circle (0.15);

    \node[below] at (4,-0.1) {$\mathcal{C}_2$};
    \draw[fill=black] (4,0) circle (0.15);

    \node[above] at (4,8) {$\mathcal{R}_2$};
    \draw[fill=blue] (4,8) circle (0.15);

    \node[above] at (-4,8) {$\mathcal{R}_1$};
    \draw[fill=blue] (-4,8) circle (0.15);
    
    \node[right] at (0.4,4) {$J^{\mathcal{E}}_{12\rightarrow 12}$};
    
    \end{tikzpicture}
    }
    \hfill
    \subfloat[\label{fig:nonlocalschematic}]{
    \begin{tikzpicture}[scale=0.7]
    
    \draw[postaction={on each segment={mid arrow}}] (-4,0) -- (-2,2) -- (-2,6) -- (-4,8);
    \draw[postaction={on each segment={mid arrow}}] (4,0) -- (2,2) -- (2,6) -- (4,8);
    \draw[postaction={on each segment={mid arrow}}] (-2,2) -- (0,4) -- (2,6);
    \draw[postaction={on each segment={mid arrow}}] (2,2) -- (0,4) -- (-2,6);
    
    \draw[dashed] (2,2) -- (0,0) -- (-2,2);
    \node[below] at (0,0) {$\ket{\Psi^+}$};
    
    \draw[fill=yellow] (-2,2) circle (0.3);
    \node[left] at (-2.3,2) {$\hat{\mathcal{V}}_1$};
    
    \draw[fill=yellow] (2,2) circle (0.3);
    \node[right] at (2.3,2) {$\hat{\mathcal{V}}_2$};
    
    \draw[fill=yellow] (-2,6) circle (0.3);
    \draw[fill=yellow] (2,6) circle (0.3);
    
    \node[below] at (-4,-0.1) {$\hat{\mathcal{C}}_1$};
    \draw[fill=black] (-4,0) circle (0.15);

    \node[below] at (4,-0.1) {$\hat{\mathcal{C}}_2$};
    \draw[fill=black] (4,0) circle (0.15);

    \node[above] at (4,8) {$\hat{\mathcal{R}}_2$};
    \draw[fill=blue] (4,8) circle (0.15);

    \node[above] at (-4,8) {$\hat{\mathcal{R}}_1$};
    \draw[fill=blue] (-4,8) circle (0.15);

    \end{tikzpicture}
    }
    \caption{Bulk and boundary perspectives on the $\mathbf{\hat{B}}_{84}$ task. (a) Causal features present in the bulk geometry. Signals from $\mathcal{C}_1$ and $\mathcal{C}_2$ may meet in the scattering region $J^{\mathcal{E}}_{12\rightarrow 12}$, then travel to either $\mathcal{R}_1$ or $\mathcal{R}_2$. The scattering region is a resource, useful for completing quantum tasks. (b) Causal features present in the boundary geometry, which lacks an entanglement scattering region. $\hat{\mathcal{C}}_1$ may send signals to $\hat{\mathcal{R}}_1$ and $\hat{\mathcal{R}}_2$, and $\hat{\mathcal{C}}_2$ may send signals to $\hat{\mathcal{R}}_1$ and $\hat{\mathcal{R}}_2$. The boundary replaces the resource of a non-empty scattering region with entanglement between $\hat{\mathcal{V}}_1$ and $\hat{\mathcal{V}}_2$. }
    \label{fig:casualschematic}
\end{figure}
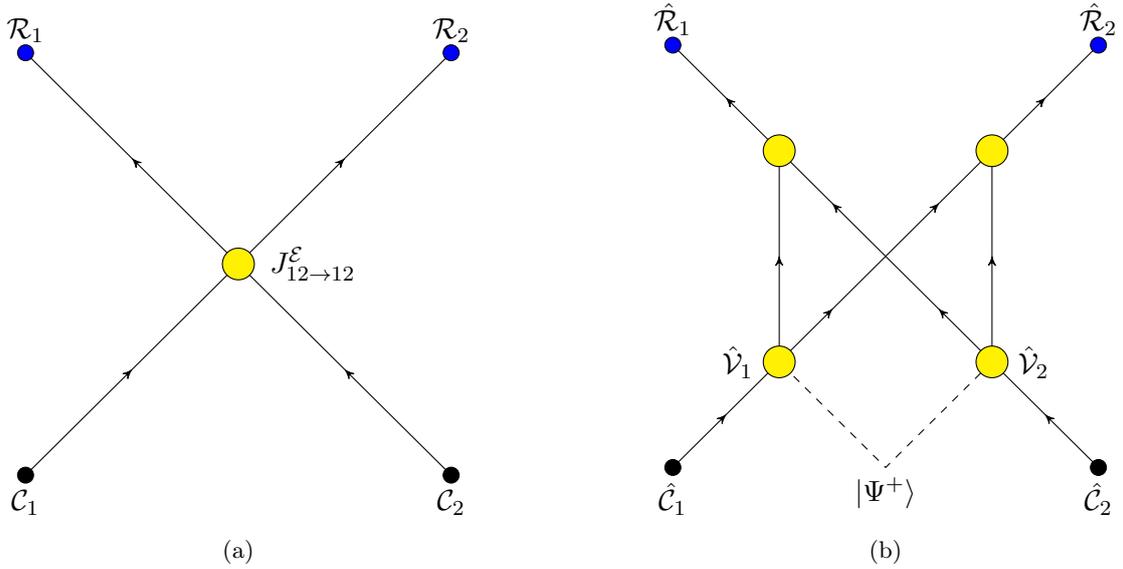

We will be interested in two strategies for completing the task: a local strategy and a non-local strategy. The local strategy is one which makes use of the scattering region
\begin{align}
    J^+(\mathcal{C}_1)\cap J^+(\mathcal{C}_2) \cap J^-(\mathcal{R}_1) \cap J^-(\mathcal{R}_2)
\end{align}
and so is only available when this region is non-empty. The non-local strategy does not use this region. The two strategies are shown in figures \ref{fig:localschematic} and \ref{fig:nonlocalschematic} respectively. We treat each below. Note that in arguing for Theorem \ref{thm:main}, we will be interested in the case where in the bulk the scattering region defined above is available and so the local strategy can be used, while in the boundary the corresponding scattering region is empty, so it is necessary to use a non-local strategy.

\subsubsection*{Local strategy for \texorpdfstring{$\mathbf{{B}}_{84}^{\times n}$}{TEXT} task}

The causal features of the task in the local strategy are captured by figure \ref{fig:localschematic}. In this case, there is a protocol which completes the task with high probability. In particular, Alice should bring $H^q\ket{b}$ from $\mathcal{C}_1$ and $q$ from $\mathcal{C}_2$ together inside the above causally defined region. Then, she applies $H^q$ to obtain $(H^q)^2\ket{b}=\ket{b}$, measures in the computational basis to learn $b$, and then sends $b$ to both $\mathcal{R}_1$ and $\mathcal{R}_2$. Assuming this can be carried out as described this completes the task with probability $p_{suc}=1$. More physically we will allow for the presence of noise in carrying out this protocol, and say the success probability satisfies $p_{suc}(\mathbf{\hat{B}}_{84}) \geq 1-\epsilon$.  

Consider repeating the $\textbf{B}_{84}$ task $n$ times in parallel. Call this repeated task $\textbf{B}_{84}^{\times n}$. This repeated task has inputs $A_1=\bigotimes_{i=1}^n H^{q_i}\ket{b_i}$ , $A_2=\{q_i\}_{i}$, and required outputs $\{b_i\}_i$ at both $\mathcal{R}_1$ and $\mathcal{R}_2$. The $q_i$ and $b_i$ are random and independent. We will declare the task to be completed successfully if a fraction $1-2\epsilon$ of the $n$ tasks are completed successfully.\footnote{This should be contrasted to the condition considered in \cite{may2019quantum,may2020holographic}, where $\mathbf{B}^{\times n}_{84}$ was declared successful only if all $n$ of the individual $\textbf{B}_{84}$ tasks were completed successfully. Taking this more relaxed condition is what eventually leads us to the bound \ref{eq:mutualinfobound}, which improves on the earlier bound.} If each task is completed with probability $p_{suc}=1-\epsilon$, then this occurs with probability
\begin{align}
    p_{suc}(\textbf{B}^{\times n}_{84}) = 1 - 2 \epsilon^{2+n}.
\end{align}
This is the success probability for the $\textbf{B}^{\times n}_{84}$ when a local strategy is available.

\subsubsection*{Non-local strategy for $\mathbf{\hat{B}}_{84}^{\times n}$ task}

In the non-local case, where no scattering region is available, we will be interested in strategies of the form shown in figure \ref{fig:nonlocalschematic}. These do not use the scattering region but instead use entanglement shared between the \emph{decision} regions
\begin{align}
    \hat{\mathcal{V}}_1 &\equiv \hat{J}^+(\hat{\mathcal{C}}_1) \cap \hat{J}^-(\hat{\mathcal{R}}_1) \cap \hat{J}^-(\hat{\mathcal{R}}_2), \nonumber \\
    \hat{\mathcal{V}}_2 &\equiv \hat{J}^+(\hat{\mathcal{C}}_2) \cap \hat{J}^-(\hat{\mathcal{R}}_1) \cap \hat{J}^-(\hat{\mathcal{R}}_2),
\end{align}
to complete the task. The regions $\hat{\mathcal{V}}_1$ and $\hat{\mathcal{V}}_2$ are relevant because they are the largest regions which contain the inputs $A_1$ and $A_2$ but are also in the past of both output regions. Rather than discussing specific strategies for completing the task non-locally, we are interested in lower bounding the amount of entanglement necessary to complete the task with high probability when using any non-local strategy.

We begin by assuming $I(\hat{\mathcal{V}}_1:\hat{\mathcal{V}}_2)=0$, that $\hat{\mathcal{C}}_i \subseteq \hat{\mathcal{V}}_i$, and considering a single instance of the task, $\mathbf{\hat{B}}^{\times 1}_{84}$. We will see that this leads to a success probability bounded strictly below 1.
\begin{lemma}\label{lemma:sucbound}
Consider the $\mathbf{\hat{B}}^{\times 1}_{84}$ task with $I(\hat{\mathcal{V}}_1:\hat{\mathcal{V}}_2)=0$. Then any strategy for completing the task has $p_{suc}(\mathbf{\hat{B}}^{\times 1}_{84})\leq \cos^2 (\pi/8)$.
\end{lemma}
This lemma is proven in \cite{tomamichel2013monogamy}. To understand it heuristically we can reason as follows. In region $\hat{\mathcal{C}}_1$ Alice holds one of the states $H^q\ket{b}$ for $q,b\in\{0,1\}$. If $\hat{\mathcal{C}}_1$ also held the basis information, Alice could measure in the computational basis $\{\ket{0},\ket{1}\}$ if $q=0$ or the Hadamard basis $\{\ket{+},\ket{-}\}$ if $q=1$ to determine $b$. Since $I(\hat{\mathcal{V}}_1:\hat{\mathcal{V}}_2)=0$ however, and $q$ is held in $\hat{\mathcal{V}}_2$, Alice must act in a way independent of $q$. Doing so, she cannot determine $b$ perfectly. Instead her optimal strategy is to measure in an intermediate basis $\{\ket{\psi_0},\ket{\psi_1}\}$ where
\begin{align}
    \ket{\psi_0} &= \cos\left( \frac{\pi}{8}\right) \ket{0} + \sin \left( \frac{\pi}{8}\right)\ket{1}, \nonumber \\ 
    \ket{\psi_1} &= \cos\left(\frac{5\pi}{8}\right) \ket{0} + \sin \left( \frac{5\pi}{8}\right)\ket{1}.
\end{align}
This leads to the $\cos^2(\pi/8)$ success probability, so that the bound in Lemma \ref{lemma:sucbound} is actually tight. Note that the assumption $\hat{\mathcal{C}}_i\subseteq \hat{\mathcal{V}}_i$ is important for this reasoning to hold. In particular, we used that $q\in \hat{\mathcal{C}}_2 \subseteq \hat{\mathcal{V}}_2$ in saying $I(\hat{\mathcal{V}}_1:\hat{\mathcal{V}}_2)=0$ implies the measurement on $A_1\in \hat{\mathcal{C}}_1\subseteq \hat{\mathcal{V}}_1$ is independent of $q$.

Next we consider the parallel repetition task, $\mathbf{\hat{B}}^{\times n}_{84}$. In \cite{tomamichel2013monogamy} the following statement has been proven.
\begin{lemma}\label{lemma:sucboundparallel}
Consider the $\mathbf{\hat{B}}^{\times n}_{84}$ task with $I(\hat{\mathcal{V}}_1:\hat{\mathcal{V}}_2)=0$, assume that $\hat{\mathcal{C}}_i\subseteq\hat{\mathcal{V}}_i$ and that the scattering region is empty. Require that a fraction $1-\delta$ of the individual $\mathbf{\hat{B}}_{84}$ tasks are successful. Then any strategy for completing the task has 
\begin{align}
    p_{suc}(\textbf{B}_{84}^{\times n})\leq \left(2^{h(\delta)}\cos^2\left( \frac{\pi}{8}\right) \right)^n \equiv \left(2^{h(\delta)}\beta\right)^n,
\end{align}
where $h(\delta)$ is the binary entropy function $h(\delta)\equiv -\delta\log_2 \delta - (1-\delta)\log_2(1-\delta)$ and the second equality defines $\beta$.
\end{lemma}
For small enough $\delta$ we have that $2^{h(\delta)}\beta<1$, so this gives a good bound on the success probability. 

In the boundary, where the scattering region is empty, the success probability is exponentially small when the mutual information is zero. Comparing to the bulk where the success probability is exponentially close to one suggests the true boundary state contains large mutual information. This is indeed the case, as we show in the next lemma. 
\begin{lemma}\label{lemma:mutualinfobound} Suppose the $\mathbf{\hat{B}}_{84}$ task is completed with probability $p_{suc}\geq 1-2 \epsilon^{2+n}$, and that the scattering region is empty. Then
\begin{align}\label{eq:mutualinfobound}
    \frac{1}{2}I(\hat{\mathcal{V}}_1:\hat{\mathcal{V}}_2)\geq n(-\log 2^{h(2\epsilon)}\beta) - 1 +O((\epsilon/\beta)^n)
\end{align}
\end{lemma}
This is proven in appendix \ref{sec:mutualinfobound}. In the next section we employ this bound to argue for the connected wedge theorem. Note that this improves on the bound presented in \cite{may2020holographic}.  

\subsection{Connected wedge theorem from quantum tasks}

In this section we give the quantum tasks argument for the connected wedge theorem. We first recall the theorem for convenience.\\

\vspace{0.2cm}
\noindent \textbf{Theorem \ref{thm:main}} \emph{\textbf{(Connected wedge theorem)} Pick four regions $\hat{\mathcal{C}}_1,\hat{\mathcal{C}}_2,\hat{\mathcal{R}}_1,\hat{\mathcal{R}}_2$ on the boundary of an asymptotically AdS spacetime. From these, define the decision regions
\begin{align}
    \hat{\mathcal{V}}_1 &\equiv \hat{J}^+(\hat{\mathcal{C}}_1) \cap \hat{J}^-(\hat{\mathcal{R}}_1) \cap \hat{J}^-(\hat{\mathcal{R}}_2) , \nonumber \\
    \hat{\mathcal{V}}_2 &\equiv \hat{J}^+(\hat{\mathcal{C}}_2) \cap \hat{J}^-(\hat{\mathcal{R}}_1) \cap \hat{J}^-(\hat{\mathcal{R}}_2).
\end{align}
Assume that $\hat{\mathcal{C}}_i \subseteq \hat{\mathcal{V}}_i$. Define the entanglement scattering region
\begin{align}
    J_{12\rightarrow 12}^{\mathcal{E}} \equiv J^+(\mathcal{C}_1)\cap J^+(\mathcal{C}_2) \cap J^-(\mathcal{R}_1) \cap J^-(\mathcal{R}_2),
\end{align}
where $\mathcal{C}_i = \mathcal{E}_W(\hat{\mathcal{C}}_i)$ and $\mathcal{R}_i = \mathcal{E}_W(\hat{\mathcal{R}}_i)$. Then, $J_{12\rightarrow 12}^{\mathcal{E}}\neq \varnothing$ implies that $\mathcal{E}_W(\hat{\mathcal{V}}_1\cup \hat{\mathcal{V}}_2)$ is connected. }\\
\vspace{0.2cm}

\noindent \begin{argument} Consider two cases. First, supposed that $\hat{\mathcal{V}}_1\cap \hat{\mathcal{V}}_2 \neq \varnothing$. Then we immediately have that the entanglement wedge of $\hat{\mathcal{V}}_1\cup \hat{\mathcal{V}}_2$ is connected, and we are done. 

Next, assume that $\hat{\mathcal{V}}_1\cap \hat{\mathcal{V}}_2 = \varnothing$. This is just the statement that the boundary scattering region
\begin{align}
    \hat{J}_{12\rightarrow 12} = \hat{J}^+(\hat{\mathcal{C}}_1)\cap \hat{J}^+(\hat{\mathcal{C}}_2) \cap \hat{J}^-(\hat{\mathcal{R}}_1) \cap \hat{J}^-(\hat{\mathcal{R}}_2)
\end{align}
is empty. By assumption however, the bulk entanglement scattering region $J_{12\rightarrow 12}^{\mathcal{E}}$ is non-empty. This implies the existence of four points $c_1,c_2,r_1,r_2$ such that 
\begin{align}
    J^+(c_1)\cap J^+(c_2) \cap J^-(r_1) \cap J^-(r_2) \neq \varnothing .
\end{align}
Choose a $\mathbf{{B}}_{84}^{\times n}$ task in the bulk with $c_1$, $c_2$ as input points and $r_1$, $r_2$ as output points. Then because the above region is non-empty, we can use the local strategy in the bulk, and we obtain a high success probability, 
\begin{align}
    p_{suc}(\mathbf{\hat{B}}_{84}^{\times n}) \geq 1 - 2\epsilon^{2+n}.
\end{align}
Next, we should discuss how large we can take $n$. The obstruction to taking $n$ arbitrarily large is that if we make use of too many qubits, the bulk protocol may change the geometry, deforming the geometry we are attempting to study. To avoid this we choose $n$ to be any order in $1/G_N$ less than linear. Then if each qubit carries some energy $\Delta E$, Einstein's equations dictate that the coupling to geometry is
\begin{align}
    G_{\mu\nu} = O(G_N \Delta E n).
\end{align}
Choosing $n<O(1/G_N)$ ensures that in the $G_N \rightarrow 0$ limit we have no backreaction, as needed to ensure we are studying the intended geometry.

Starting with $\mathbf{{B}}_{84}^{\times n}$, we label the corresponding boundary task by $\mathbf{\hat{B}}_{84}^{\times n}$. Following the discussion in section \ref{sec:framework}, we know $\mathbf{\hat{B}}_{84}^{\times n}$ has the same inputs and outputs as the corresponding bulk task. Further, we have by assumption that
\begin{align}
    c_i \in \mathcal{E}_W(\hat{\mathcal{C}}_i), \nonumber \\
    r_i \in \mathcal{E}_W(\hat{\mathcal{R}}_i).
\end{align}
Thus $\hat{\mathcal{C}}_i$ is an authorized region for $A_i$ in the boundary task, and $\hat{\mathcal{R}}_i$ is an authorized region for $B_i$. Since $\hat{\mathcal{C}}_i\subseteq \hat{\mathcal{V}}_i$, and by assumption the boundary scattering region $\hat{\mathcal{V}}_1\cap \hat{\mathcal{V}}_2$ is empty, the boundary uses a non-local strategy. Lemma \ref{lemma:mutualinfobound} then applies and we can conclude
\begin{align}
    \frac{1}{2}I(\hat{\mathcal{V}}_1:\hat{\mathcal{V}}_2) \geq n (-\log_2\beta ) - 1 + O((\epsilon/\beta)^n).
\end{align}
Since $n$ is any order less than $O(1/G_N)$, we can conclude that $I(\hat{\mathcal{V}}_1:\hat{\mathcal{V}}_2) = O(1/G_N)$, which occurs only when $\mathcal{E}_W(\hat{\mathcal{V}}_1\cup \hat{\mathcal{V}}_2)$ is connected.
\end{argument}

Notice that the Ryu-Takayanagi formula only appears in this argument in the last step. Without ever using the Ryu-Takayanagi formula, we can still conclude that the mutual information is order $1/G_N$. We also could keep $\epsilon$ fixed in this argument. This was possible because of the improved bound \ref{eq:mutualinfobound}. Using the earlier bound as it appeared in \cite{may2020holographic}, it was necessary to appeal to the Ryu-Takayanagi formula and to argue one can take $\epsilon\rightarrow 0$ in the $G_N\rightarrow 0$ limit. It is interesting that one can reach conclusions about boundary entanglement from bulk geometry without using the Ryu-Takayanagi formula. 

We should note that there is a gap in the argument above, which was noted also in \cite{may2020holographic}. In particular the causal diagram \ref{fig:nonlocalschematic} does not include the regions that sit between $\hat{\mathcal{V}}_1$ and $\hat{\mathcal{V}}_2$. In general, these can be made use of to complete the $\mathbf{\hat{B}}_{84}^{\times n}$ task without entanglement.\footnote{See appendix B of \cite{may2020holographic}.} However, such strategies require GHZ correlations in the CFT, which are not expected \cite{Nezami:2016zni}. As well, it seems possible to rule out such strategies by keeping Alice ignorant of the location of the regions $\hat{\mathcal{C}}_i$ before the beginning of the task, in which case she cannot coordinate actions with the intermediate regions. We leave better understanding this to future work, and for now rely on gravitational reasoning to provide a complete proof.

\section{Relativistic perspective on the connected wedge theorem}\label{sec:GRperspectiveonCWtheorem}

\subsection{Relativistic proof}\label{subsec:GRproof}

The proof of Theorem \ref{thm:main} is nearly identical to the proof of the earlier connected wedge theorem, which appears already in \cite{may2020holographic}. For readers familiar with the earlier proof, it suffices to note that the only change is to replace the causal horizon $\partial[J^-(r_1)\cap J^-(r_2)]$ with the null sheet $\partial[J^-(\mathcal{R}_1)\cap J^-(\mathcal{R}_2)]$. The key point is that $\partial[J^-(r_1)\cap J^-(r_2)]$ and $\partial[J^-(\mathcal{R}_1)\cap J^-(\mathcal{R}_2)]$ meet the boundary along the same curves, and both surfaces have area theorems for past directed null geodesics. This allows the two surfaces to play similar roles in the proof.

To be self contained, we also present the proof briefly here. Assume that the null energy condition holds. Work by contradiction by assuming that the entanglement scattering region is non-empty and the minimal area extremal surface homologous to $\hat{\mathcal{V}}_1\cup \hat{\mathcal{V}}_2$ consists of a connected component $\gamma_{{\mathcal{V}}_1}$ homologous to $\hat{\mathcal{V}}_1$ and a connected component $\gamma_{{\mathcal{V}}_2}$ homologous to $\hat{\mathcal{V}}_2$. Then, the maximin prescription \cite{maximin} for finding Ryu-Takayanagi surfaces dictates that there will exists a Cauchy slice $\Sigma$ of the bulk in which $\gamma_{{\mathcal{V}}_1}\cup \gamma_{{\mathcal{V}}_2}$ is minimal, where by $\gamma_{{X}}$ we always mean the Ryu-Takayanagi surface for a boundary region $\hat{X}$. We then construct a codimension $1$ surface we call the \emph{null membrane} $\mathcal{N}_\Sigma$, along with the codimension $2$ \emph{contradiction surface} $\mathcal{C}_\Sigma$. The null membrane facilitates a comparison of the area of $\gamma_{{\mathcal{V}}_1}\cup \gamma_{{\mathcal{V}}_2}$ and the contradiction surface. The contradiction surface will turn out to have less area than the candidate surface $\gamma_{{\mathcal{V}}_1}\cup \gamma_{{\mathcal{V}}_2}$, which provides the contradiction. 

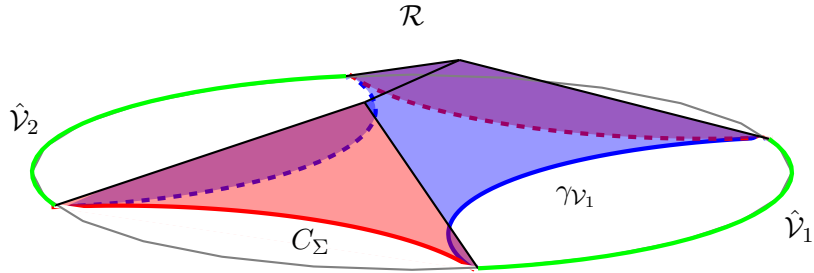
\begin{figure}
    \centering
    \tdplotsetmaincoords{15}{0}
    \begin{tikzpicture}[scale=2.5,tdplot_main_coords]
    \tdplotsetrotatedcoords{0}{30}{0}
    
    \begin{scope}[tdplot_rotated_coords]
    
    \draw[domain=40:140,variable=\x,smooth,ultra thick,blue] plot ({2*sin(\x+180)+2.53}, {0}, {2*cos(\x+180)});
    \draw[dashed,domain=40:140,variable=\x,smooth,ultra thick,blue] plot ({-(2*sin(\x+180)+2.53)}, {0}, {2*cos(\x+180)});
    
    \draw[dashed,domain=-40:40,variable=\x,smooth,ultra thick,red] plot ({-(2*sin(\x+180))}, {0}, {2*cos(\x+180)+3.1});
    \draw[domain=-40:40,variable=\x,smooth,ultra thick,red] plot ({-(2*sin(\x+180))}, {0}, {-(2*cos(\x+180)+3.1)});
    
    \fill[red,opacity=0.4,domain=40:-40] (-1.286,0,1.532) -- (0,0.5,0.5) -- (1.286,0,1.532) plot ({-(2*sin(\x+180))}, {0}, {2*cos(\x+180)+3.1});
    
    \fill[blue,opacity=0.4,domain=40:140] (1.286,0,1.532) -- (0,0.5,0.5) -- (0,0.5,-0.5) -- (1.286,0,-1.532) plot ({2*sin(\x+180)+2.53}, {0}, {2*cos(\x+180)});
    \draw[line width=1mm, white] (1.27,0,-1.5) -- (1.27,0,1.5);
    
    \fill[blue,opacity=0.4,domain=40:140] (-1.286,0,1.532) -- (0,0.5,0.5) -- (0,0.5,-0.5) -- (-1.286,0,-1.532) plot ({-(2*sin(\x+180)+2.53)}, {0}, {2*cos(\x+180)});
    \draw[line width=1mm, white] (-1.27,0,-1.5) -- (-1.27,0,1.5);
    
    \fill[red,opacity=0.4,domain=40:-40] (-1.286,0,-1.532) -- (0,0.5,-0.5) -- (1.286,0,-1.532) plot ({-(2*sin(\x+180))}, {0}, {-(2*cos(\x+180)+3.1)});
    \draw[ultra thick,white] (-1.286,0,-1.537) -- (1.286,0,-1.537);

    \draw[domain=0:360,thick,gray] plot ({2*cos(\x)},0, {2*sin(\x)});
    
    \draw[black,thick] (1.286,0,1.532) -- (0,0.5,0.5);
    \draw[black,thick] (1.286,0,-1.532) -- (0,0.5,-0.5);
    \draw[black,thick] (-1.286,0,1.532) -- (0,0.5,0.5);
    \draw[black,thick] (-1.286,0,-1.532) -- (0,0.5,-0.5);
    
    \draw[thick] (0,0.5,-0.5) -- (0,0.5,0.5);
    
    \draw [green,ultra thick,domain=40:140] plot ({2*cos(\x-90)},0, {2*sin(\x-90)});
    \draw [green,ultra thick,domain=40:140] plot ({-2*cos(\x-90)},0, {2*sin(\x-90)});
    
    \node[above] at (0,0.75,0) {$\mathcal{R}$};
    \node at (0.25,0,-1.5) {$C_\Sigma$};
    
    \node at (1,0,0) {$\mathcal{\gamma}_{\mathcal{V}_1}$};
    
    \node[right] at (2.2,0,0) {$\hat{\mathcal{V}}_1$};
    \node[left] at (-2.2,0,0) {$\hat{\mathcal{V}}_2$};

    \end{scope}
    \end{tikzpicture}
    \caption{The null membrane. The blue surface is the lift $\mathcal{L}$, which is generated by the null geodesics defined by the inward, future pointing null normals to $\gamma_{\mathcal{V}_1}\cup \gamma_{\mathcal{V}_2}$, where $\gamma_{{\mathcal{V}}_i}$ is the Ryu-Takayanagi surface for region $\hat{\mathcal{V}}_i$. The red surfaces make up the slope $\mathcal{S}_\Sigma$, which is generated by the null geodesics defined by the inward, past directed null normals to $\gamma_{{\mathcal{R}}_1}\cup \gamma_{{\mathcal{R}}_2}$. The ridge $\mathcal{R}$ is where null rays from $\gamma_{{\mathcal{V}}_1}$ and $\gamma_{{\mathcal{V}}_2}$ collide. The contradiction surface $C_\Sigma$ is where the slope meets a specified Cauchy surface $\Sigma$.}
    \label{fig:nullmembrane}
\end{figure}

The null membrane is illustrated in figure \ref{fig:nullmembrane}. It is defined as the union of two surfaces, called the \emph{lift} and the \emph{slope}. The lift is defined by
\begin{align}
    \mathcal{L}=\partial J^+(\mathcal{V}_1\cup \mathcal{V}_2) \cap J^-(\mathcal{R}_1) \cap J^-(\mathcal{R}_2).
\end{align}
The slope is defined by
\begin{align}
    \mathcal{S}_\Sigma = \partial [J^-(\mathcal{R}_1) \cup J^-(\mathcal{R}_2)] \cap J^-[\partial J^+(\mathcal{V}_1\cup \mathcal{V}_2)] \cap J^+(\Sigma).
\end{align}
The contradiction surface is defined by $\mathcal{C}_\Sigma = S_\Sigma \cap \Sigma$. Note that $\partial J^+(\mathcal{V}_1\cup \mathcal{V}_2)$ is generated by geodesics starting on the inward, future pointing, null normals to the extremal surface $\gamma_{{\mathcal{V}}_2}\cup \gamma_{{\mathcal{V}}_2}$. Similarly, $\partial [J^-(\mathcal{R}_1) \cup J^-(\mathcal{R}_2)]$ is generated by geodesics starting on the inward, past directed, null normals to $\gamma_{{\mathcal{R}}_1}\cup \gamma_{{\mathcal{R}}_2}$. 

To repeat the proof of \cite{may2020holographic} we need to establish various features of the null membrane. The first is that the area of the past directed null geodesics that generate the slope have decreasing area. This holds because this congruence is defined by beginning with the inward, past directed null normals to $\gamma_{{\mathcal{R}}_1}\cup \gamma_{{\mathcal{R}}_2}$. Since this surface is extremal, the focusing theorem (which assumes the null energy condition) implies that this congruence has decreasing area. Similarly, the null normals that generate the lift also have decreasing area.

The second needed feature of the null membrane is that the contradiction surface is homologous to $\hat{\mathcal{V}}_1 \cup \hat{\mathcal{V}}_2$. This follows because the restriction of the contradiction surface to the boundary is the spacelike boundary of $\hat{\mathcal{V}}_1\cup \hat{\mathcal{V}}_2$. 

The null membrane can be used to establish that the contradiction surface has less area than the candidate surface. To see this, consider pushing the candidate surface forward along the congruence defined by the lift, removing any generators which collide. Continue pushing the surface forward until reaching the slope. When doing so, any generators which reach the \emph{ridge} will be removed, where the ridge is defined by
\begin{align}
    \mathcal{R} = \partial J^+(\mathcal{V}_1)\cap \partial J^+(\mathcal{V}_2) \cap J^-(\mathcal{R}_1) \cap J^-(\mathcal{R}_2).
\end{align}
See also figure \ref{fig:nullmembrane}. Assume momentarily that the ridge is non-empty. Then after pushing forward the surface will consist of two disconnected components which sit on the slope. Finally, push the surface backwards along the slope until it reaches $\Sigma$, and becomes the contradiction surface. Because the null congruences defining the lift and the slope begin as normal vectors to extremal surfaces, moving into the future along the lift and into the past along the slope both decrease area. Removing colliding generators also decreases the area. Thus we can conclude the contradiction surface has less area than the candidate surface, as needed. 

To justify our assumption that the ridge is non-empty, note that this occurs whenever the entanglement scattering region $J^{\mathcal{E}}_{12\rightarrow 12}$ is non-empty, since by assumption $J^{\mathcal{E}}_{12\rightarrow 12}$ is non-empty and
\begin{align}
    J^{\mathcal{E}}_{12\rightarrow 12} \subseteq J^+(\mathcal{V}_1) \cap J^+(\mathcal{V}_2) \cap J^-(\mathcal{R}_1) \cap J^-(\mathcal{R}_2),
\end{align}
so that the region on the right is non-empty. But this region being non-empty means $J^+(\mathcal{V}_1)$ and $J^+(\mathcal{V}_2)$ must meet in the past of $\mathcal{R}_1$ and $\mathcal{R}_2$, which means the ridge is non-empty. 

\subsection{The entanglement scattering region is inside the entanglement wedge}\label{sec:scatteringinclusion}

In the context of the connected wedge theorem with input and output regions taken to be points, the authors of \cite{may2020holographic} noted that the scattering region sits inside of the entanglement wedge of $\hat{\mathcal{V}}_1\cup \hat{\mathcal{V}}_2$, at least in the context of $2+1$ bulk dimensions. We can straightforwardly adapt their argument to our context to see that the larger entanglement scattering region is also inside of the entanglement wedge, again in $2+1$ bulk dimensions. 

To see this, define the region
\begin{align}
    X = \overline{J^+[\hat{\mathcal{V}}_1 \cup \hat{\mathcal{V}}_2]^c \cap [\hat{\mathcal{V}}_1 \cup \hat{\mathcal{V}}_2]^c}.
\end{align}
This is the closure of the spacelike complement of $\hat{\mathcal{V}}_1\cup \hat{\mathcal{V}}_2$. In $2+1$ dimensions, this consists of the domains of dependence of two intervals which we call $\hat{X}_1$ and $\hat{X}_2$. Note that $\hat{\mathcal{V}}_1\cup \hat{\mathcal{V}}_2 \cup \hat{X}_1 \cup \hat{X}_2$ is a complete Cauchy slice of the boundary, which we extend into the bulk to some Cauchy slice $\Sigma$. We will choose this extension such that $\gamma_{X_1}$ and $\gamma_{X_2}$ are contained in $\Sigma$.

Next we note that $\hat{\mathcal{R}}_1$ is inside the domain of dependence of $\hat{\mathcal{V}}_1\cup \hat{X}_1 \cup \hat{\mathcal{V}_2}$, which we label $\hat{D}_1$, while $\hat{\mathcal{R}}_2$ sits inside the domain of dependence $\hat{\mathcal{V}}_1\cup \hat{X}_2 \cup \hat{\mathcal{V}_2}$, which we label $\hat{D}_2$. Because of this, $J^-(\mathcal{R}_1)$ will be inside $J^-(D_1)$, while $J^-(\mathcal{R}_2)$ will be inside $J^-(D_2)$. Consequently we learn
\begin{align}\label{eq:pastoffutureboundary}
    J^-(\mathcal{R}_1)\cap J^-(\mathcal{R}_2) \subseteq J^-(D_1) \cap J^-(D_2).
\end{align}
Notice that, assuming the entanglement wedge $\mathcal{E}_W(\hat{\mathcal{V}}_1\cup \hat{\mathcal{V}}_2)$ is connected, the future boundary of $J^-(D_1) \cap J^-(D_2)$ is also the future boundary of $\mathcal{E}_W(\hat{\mathcal{V}}_1\cup \hat{\mathcal{V}}_2)$. This is because the entangling surface for $\mathcal{E}_W(\hat{\mathcal{V}}_1\cup \hat{\mathcal{V}}_2)$ consists of two components, one of which is homologous to $\hat{X}_1$ and the other homologous to $\hat{X}_2$, and so these two components are the entangling surfaces for $\hat{D}_1$ and $\hat{D}_2$ respectively. Thus equation \ref{eq:pastoffutureboundary} gives that $J^-(\mathcal{R}_1)\cap J^-(\mathcal{R}_2)$ is in the past of the future boundary of $\mathcal{E}_W(\hat{\mathcal{V}}_1 \cup \hat{\mathcal{V}}_2)$. Since the scattering region is a subregion of $J^-(\mathcal{R}_1)\cap J^-(\mathcal{R}_2)$, it follows that this also holds for the scattering region.

It remains to show that the scattering region is to the future of the past boundary of $\mathcal{E}_W(\hat{\mathcal{V}}_1\cup \hat{\mathcal{V}}_2)$. This is immediate, because the past of $\gamma_{X_1}$ and $\gamma_{X_2}$ meets the boundary along the past boundaries of $\mathcal{V}_1$ and $\mathcal{V}_2$. Thus any points in the future of $\mathcal{V}_1$ and $\mathcal{V}_2$ must be in the future of this past boundary.   

\section{Discussion}\label{sec:discussion}

In this article we have expanded the holographic quantum tasks framework to include inputs and outputs encoded into arbitrary access structures. We've illustrated the usefulness of this framework by using this construction to motivate the improved connected wedge theorem, which we could then verify using a geometric proof.

\subsection{Non-triviality of the theorem in various spacetimes}\label{sec:non-trivial}

We have noted that if the boundary regions $\hat{\mathcal{V}}_1$ and $\hat{\mathcal{V}}_2$ overlap, then the conclusion of the connected wedge theorem is trivial, since in that case $\hat{\mathcal{V}}_1$ and $\hat{\mathcal{V}}_2$ immediately have a connected entanglement wedge. Non-trivial configurations in global AdS$_{2+1}$ exist and are discussed explicitly in \cite{may2019quantum,may2020holographic}, while \cite{may2020holographic} also noted non-trivial configurations exist in the AdS soliton, and here we have given non-trivial configurations for Poincar\'e-AdS$_{2+1}$ (which only exist when considering the region based statement).  

We have not explored in detail however if the region based connected wedge theorem applies non-trivially in higher dimensions. In Poincar\'e-AdS$_{3+1}$ it seems straightforward to construct non-trivial configurations by defining the input and output regions to be as defined in section \ref{sec:planar}, but extended infinitely in the extra transverse direction. In global-AdS$_{3+1}$ it is less clear how to construct such non-trivial configurations, though one plausible avenue is to begin with the Poincar\'e configurations and consider their embedding into the global spacetime. We leave understanding this in detail to future work.

\subsection{Improved bounds on the mutual information}

One technical improvement over \cite{may2020holographic} made in this work is a more robust handling of possible noise in the bulk protocol. In particular we proved
\begin{align}
    \frac{1}{2}I(\hat{\mathcal{V}}_1:\hat{\mathcal{V}}_2) \geq n (-\log_2\beta ) - 1 + O((\epsilon/\beta)^n),
\end{align}
where $\epsilon$ was the error in completing a single instance of the $\textbf{B}_{84}$ task. Since we argued $n$ can be taken to be any order less than $O(1/G_N)$, this bound allows us to directly conclude that the mutual information between two regions $\hat{\mathcal{V}}_1$ and $\hat{\mathcal{V}}_2$ which have a scattering region is $O(1/G_N)$. In \cite{may2020holographic}, using a weaker bound, it was only possible to prove the mutual information was $O(1/G_N)$ by first using the HRT formula to see that the mutual information is either $O(1/G_N)$ or $O(1)$. \\

\subsection{Towards a causal structure-entanglement theorem with a converse}

From a tasks perspective the failure of Theorem \ref{thm:main} to have a converse is tied to the fact that we are interested in a fixed bulk geometry. While protocols that take place in that fixed geometric background have a boundary description, many boundary protocols will deform the geometry. Because of this, the bulk and boundary success probabilities are related by an inequality, $p_{suc}(\mathbf{T})\leq p_{suc}(\mathbf{\hat{T}})$, so that sufficient entanglement to do the task in the boundary does not imply the task can be done in the bulk fixed geometry, and so does not signal the appearance of a bulk scattering region.

One interesting possibility is that large boundary correlation when measured in some way other than the mutual information will imply the existence of a bulk entanglement scattering region. In particular, this hypothetical measure of correlation should count only entanglement that can be made use of by operations that preserve the bulk geometry. Then, its appearance would signal that there should be a bulk protocol in that geometry which completes the required task, which in turn would imply the existence of the scattering region.\footnote{We thank Jon Sorce for discussion on these points.} \\

\noindent \textbf{Acknowledgements}\vspace{0.3cm}

\noindent I thank Kfir Dolev, Jon Sorce, and Jason Pollack for valuable discussions and feedback on drafts of this article. I am supported by a C-GSM award given by the National Science and Engineering Research Council of Canada.

\appendix 

\section{Lower bound on mutual information from success probability}\label{sec:mutualinfobound}

In this section we prove the lower bound on mutual information \ref{eq:mutualinfobound}. Our starting point is Lemma \ref{lemma:sucboundparallel} which bounds the success probability for states with zero mutual information. Our argument will show that states with high probability must be far from these zero probability states in terms of trace distance, which we can then translate to a bound on mutual information. Note that the discussion here is a repetition of an argument in \cite{may2020holographic}, but, because various parameters are changed in our context, we've included the proof with updated parameters here. 

We begin by recalling a continuity bound on success probability for any quantum task, stated earlier in \cite{may2020holographic}. 
\begin{lemma}\label{lemma:continuity}
Consider a quantum task which takes as input a quantum system $A$. Then the probability of completing the task, call it $p_{suc}$, satisfies the continuity bound
\begin{align}
    |p_{suc}(\rho_A)-p_{suc}(\sigma_A)| \leq \frac{1}{2}|| \rho_A-\sigma_A||_1.
\end{align}
\end{lemma}
The proof follows by viewing the task as a procedure for distinguishing $\rho$ from $\sigma$. The maximal success probability of distinguishing states can be written in terms of the trace distance, which leads to the above inequality. Intuitively, we should understand the lemma as saying that nearby states produce nearby success probabilities. 

For the task $\mathbf{\hat{B}}^{\times n}_{84}$, we have
\begin{align}\label{eq:psucbounds}
    p_{suc}(\rho_{\hat{\mathcal{V}}_1\hat{\mathcal{V}}_2}) &\geq 1 - 2\epsilon^{2+n}, \nonumber \\
    p_{suc}(\rho_{\hat{\mathcal{V}}_1}\otimes \rho_{\hat{\mathcal{V}}_2} ) &\leq (2^{h(\delta)}\beta)^n.
\end{align}
The state $\rho_{\hat{\mathcal{V}}_1\hat{\mathcal{V}}_2}$ is any boundary state where the bulk scattering region is non-empty, while $\rho_{\hat{\mathcal{V}}_1}\otimes \rho_{\hat{\mathcal{V}}_2}$ is the tensor product of its marginals. The second bound follows from Lemma \ref{lemma:sucboundparallel}.

The remainder of the argument consists of relating the trace distance to the relative entropy. In particular we have that the trace distance and fidelity are related by \cite{fuchs1999cryptographic,wilde2013quantum}
\begin{align}\label{eq:tracedistanceandfidelity}
    \frac{1}{2}||\rho-\sigma||_1 \leq \sqrt{1-F(\rho,\sigma)}
\end{align}
where $F(\rho,\sigma)$ is the fidelity. Additionally, 
\begin{align}\label{eq:fidelityandrelativeentropy}
    -2\log F(\rho,\sigma) \leq D(\rho||\sigma)
\end{align}
where $D(\rho|\sigma)$ is the relative entropy. The final observation is that
\begin{align}
    D(\rho_{AB}||\rho_A\otimes \rho_B) = I(A:B)_\rho 
\end{align}
Combining inequalities \ref{eq:tracedistanceandfidelity} and \ref{eq:fidelityandrelativeentropy} and Lemma \ref{lemma:continuity} to lower bound the relative entropy, and hence the mutual information, in terms of success probabilities we find
\begin{align}
    I(\hat{\mathcal{V}}_1:\hat{\mathcal{V}}_2)_{\rho} \geq -2\log[1-|p_{suc}(\rho_A)-p_{suc}(\rho_{AB}\otimes \rho_B)|^2].
\end{align}
Finally using our bounds on success probability \ref{eq:psucbounds} we obtain
\begin{align}
    \frac{1}{2}I(\hat{\mathcal{V}}_1:\hat{\mathcal{V}}_2)\geq n(-\log 2^{h(2\epsilon)}\beta) - 1 +O((\epsilon/\beta)^n)
\end{align}
as claimed.

\section{Non-trivial configurations in Poincar\'e-AdS\texorpdfstring{$_{2+1}$}{TEXT}}\label{sec:poincareexamples}

\begin{figure}
    \centering
    \begin{tikzpicture}[scale=0.7]
    
    \draw[thick, black, fill=black!60!,opacity=0.8] (-6,0) -- (-4,2) -- (-2,0) -- (-4,-2) -- (-6,0);
    
    \draw[thick, black, fill=black!60!,opacity=0.8] (6,0) -- (4,2) -- (2,0) -- (4,-2) -- (6,0);
    
    \draw[lightgray] (-10,8) -- (10,8) -- (10,-2) -- (-10,-2) -- (-10,8);
    
    \draw[fill=black] (-4,-2) circle (0.15);
    \draw[fill=black] (4,-2) circle (0.15);
    \node[below left] at (-4,-2) {$c_1$};
    \node[below right] at (4,-2) {$c_2$};

    \draw[thick,fill=blue,opacity=0.3] (-4,2) -- (0,6) -- (4,2) -- (0,-2) -- (-4,2);
    \draw[thick,fill=blue,opacity=0.3] (-4,2) -- (-10,8) -- (-10,-2) -- (-8,-2) -- (-4,2);
    \draw[thick,fill=blue,opacity=0.3] (4,2) -- (10,8) -- (10,-2) -- (8,-2) -- (4,2);
    
    \node at (4,0) {$\hat{\mathcal{V}}_2$};
    \node at (-4,0) {$\hat{\mathcal{V}}_1$};

    \node at (0,2) {$\hat{\mathcal{R}}_1$};
    \node at (-8,2) {$\hat{\mathcal{R}}_2$};
    \node at (8,2) {$\hat{\mathcal{R}}_2$};
    
    \draw[fill=black] (0,6) circle (0.15);
    \node[above right] at (0,6) {$r_1$};
    \draw[fill=black] (0,-2) circle (0.15);
    \node[below right] at (0,-2) {$e_1$};
    
    \draw[fill=black] (-6,0) circle (0.15);
    \node[above left] at (-6,0) {$q_{1L}$};
    \draw[fill=black] (-2,0) circle (0.15);
    \node[above right] at (-2,0) {$q_{1R}$};
    
    \draw[fill=black] (6,0) circle (0.15);
    \node[above right] at (6,0) {$q_{2R}$};
    \draw[fill=black] (2,0) circle (0.15);
    \node[above left] at (2,0) {$q_{2L}$};
    
    \end{tikzpicture}
    \caption{A typical choice of regions $\hat{\mathcal{C}}_1=\hat{\mathcal{V}}_1$, $\hat{\mathcal{C}}_2=\hat{\mathcal{V}}_2$, $\hat{\mathcal{R}}_1, \hat{\mathcal{R}}_2$ in the boundary of Poincar\'e-AdS which leads to a non-trivial conclusion in the connected wedge Theorem \ref{thm:main}. The regions are conveniently specified by choosing the four points $c_1,c_2,r_1,e_1$. }
    \label{fig:poincaresetup}
\end{figure}
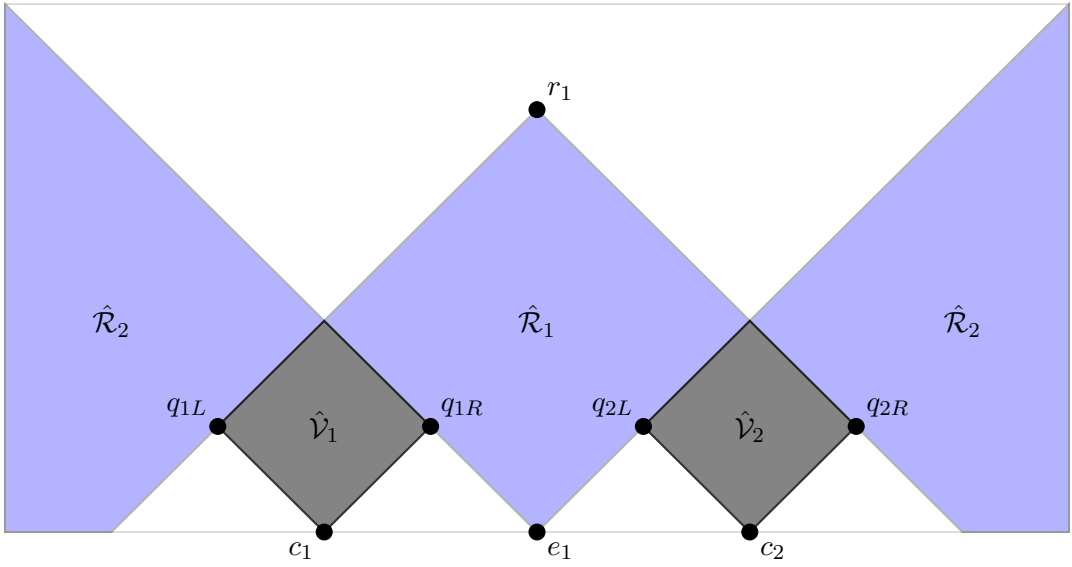

Here we explicitly check there are non-trivial configurations for the connected wedge theorem in (pure) Poincar\'e-AdS. 

Begin by considering points at\footnote{Here and throughout the section coordinates are labelled by $(t,x)$.}
\begin{align}
    c_1 &= (0,-\ell), \nonumber  \\
    c_2 &= (0,\ell), \nonumber \\
    r_1 &= (T_r,0), \nonumber \\
    e_1 &= (T_e,0).
\end{align}
The set-up is shown in figure \ref{fig:poincaresetup}. This defines regions,
\begin{align}
    \hat{\mathcal{R}}_1 &= \hat{J}^+(e_1) \cap \hat{J}^-(r_1), \nonumber \\
    \hat{\mathcal{V}}_1 &= [\hat{J}^+(c_1) \cap \hat{J}^-(r_1)] \setminus \hat{J}^+(e_1), \nonumber \\
    \hat{\mathcal{V}}_2 &= [\hat{J}^+(c_2) \cap \hat{J}^-(r_1)] \setminus \hat{J}^+(e_1).
\end{align}
We then define $\hat{\mathcal{R}}_2$ as the set of points spacelike separated from $\hat{\mathcal{R}}_1$. 

We are interested in finding choices of points such that the entanglement scattering region is non-empty. Because we are in pure AdS, we find
\begin{align}
    J^+(\mathcal{V}_1) &= J^+(c_1), \nonumber \\
    J^+(\mathcal{V}_2) &= J^+(c_2), \nonumber \\
    J^-(\mathcal{R}_1) &= J^-(r_1), \nonumber \\
    J^-(\mathcal{R}_2) &= [J^+(e_1)]^c.
\end{align}
Thus a non-empty entanglement scattering region amounts to configurations such that 
\begin{align}
    J^{\mathcal{E}}_{12\rightarrow 12} = J^+(c_1)\cap J^+(c_2) \cap J^-(r_1)\cap [J^+(e_1)]^c
\end{align}
is non-empty. These light cones are given in Poincar\'e-AdS by
\begin{align}
    J^+(c_1) &= \{(t,x,z):(x+\ell)^2+z^2-t^2\leq 0 \}, \\
    J^+(c_2) &= \{(t,x,z):(x-\ell)^2+z^2-t^2\leq 0 \}, \\
    J^+(r_1) &= \{(t,x,z):x^2+z^2-(T_r-t)^2\leq 0 \}, \\
    [J^+(e_1)]^c &= \{(t,x,z):x^2+z^2-(T_e-t)^2\geq 0 \}.
\end{align}
By symmetry, if the scattering region is non-empty it will include at least one point with $x=0$. Thus it suffices to find a solution to
\begin{align}
    \ell^2+z^2-t^2 &\leq 0, \\
    z^2-(T_r-t)^2 &\leq 0, \\
    z^2 - (T_e-t)^2 &\geq 0.
\end{align}
Eliminating $z$ and $t$ from these inequalities we find one constraint,
\begin{align}\label{eq:poincarecondition}
    \boxed{T_eT_r-\ell^2 > 0}.
\end{align}
We need to check this can be satisfied while also having $\hat{\mathcal{V}}_1\cap \hat{\mathcal{V}}_2=\varnothing$. In our setup this occurs when the rightmost point in $\hat{\mathcal{V}}_1$ is to the left of $x=0$ and the leftmost point in $\hat{\mathcal{V}}_2$ to the right of $x=0$. Via some Minkowski space geometry we can work out the end-points of the regions $\hat{\mathcal{V}}_1,\hat{\mathcal{V}}_2$,
\begin{align}\label{eq:poincarepoints}
    q_{1L} &= \left(\frac{T_r-\ell}{2},-\frac{T_r+\ell}{2}\right), \nonumber \\
    q_{1R} &= \left(\frac{T_e+\ell}{2},\frac{T_e-\ell}{2}\right), \nonumber \\
    q_{2R} &= \left(\frac{T_r-\ell}{2},\frac{T_r+\ell}{2}\right),\nonumber \\
    q_{2L} &= \left( \frac{T_e+\ell}{2},\frac{\ell-T_e}{2}\right),
\end{align}
where $q_{iL}$ is the left spacelike boundary of $\hat{\mathcal{V}}_i$, and $q_{iR}$ is the right spacelike boundary of $\hat{\mathcal{V}}_i$ (see figure \ref{fig:poincaresetup}). We see that keeping $\hat{\mathcal{V}}_1$ and $\hat{\mathcal{V}}_2$ disjoint just requires $T_e<\ell$. Given a choice of $T_e,\ell$ such that $T_e<\ell$, we can always choose $T_r$ sufficiently large to satisfy \ref{eq:poincarecondition}, guaranteeing the existence of the scattering region. This establishes that there are non-trivial configurations for Theorem \ref{thm:main} in Poincar\'e-AdS$_{2+1}$.

Given that the scattering region exists, Theorem \ref{thm:main} concludes that $\hat{\mathcal{V}}_1$ and $\hat{\mathcal{V}}_2$ should have a connected entanglement wedge. We can also check this explicitly in the simple setting considered here. In particular we would like to understand when
\begin{align}\label{eq:areacondition}
    \mathcal{A}_{min}[(q_{1L},q_{1R})]+\mathcal{A}_{min}[(q_{2L},q_{2R})] \geq \mathcal{A}_{min}[(q_{1L},q_{2R})] + \mathcal{A}_{min}[(q_{1R},q_{2L})].
\end{align}
To calculate the minimal area surface for one of the $\hat{\mathcal{V}}_i$, we need the invariant length of its cross section. This is straightforward to work out from \ref{eq:poincarepoints}, 
\begin{align}
    \Delta x^2 - \Delta t^2 = \ell(T_r-T_e).
\end{align}
The intervals $(q_{1L},q_{2R})$ and $(q_{1R},q_{2L})$ are on a constant $t$ surface, so we can just use their widths. The areas of the minimal surfaces are then
\begin{align}
    \mathcal{A}_{min}[(q_{iL},q_{iR})] &= 2L_{AdS}\ln \left( \frac{\sqrt{\ell(T_r-T_e)}}{\epsilon}\right), \nonumber \\
    \mathcal{A}_{min}[(q_{1L},q_{2R})] &=  2L_{AdS}\ln \left( \frac{{\ell-T_e}}{\epsilon}\right), \nonumber \\
    \mathcal{A}_{min}[(q_{1R},q_{2L})] &= 2L_{AdS}\ln \left( \frac{T_r+\ell}{\epsilon}\right). 
\end{align}
The condition \ref{eq:areacondition} reduces then to just 
\begin{align}
    \boxed{T_eT_r-\ell^2 \geq 0}
\end{align}
which is the same condition we found the existence of the scattering region, verifying the theorem explicitly in this simple case.

\bibliographystyle{jhep}
\bibliography{biblio}

\end{document}